\documentclass[a4paper,11pt]{article}
\pdfoutput=1 %

\usepackage{styles/jheppub} %
\usepackage[T1]{fontenc} %

\usepackage{styles/main}
\usepackage{styles/defs}

\graphicspath{{assets/}}

\makeatletter
\gdef\@fpheader{}
\makeatother

\title{\Yadism: Yet Another Deep-Inelastic Scattering Module}

\author[a,b]{Alessandro Candido}
\author[a,c,d]{Felix Hekhorn}
\author[e,f]{Giacomo Magni}
\author[e,f]{Tanjona R. Rabemananjara}
\author[g]{Roy Stegeman}

\affiliation[a]{
  Tif Lab, Dipartimento di Fisica, Universit\`a di Milano and\\
  INFN, Sezione di Milano, Via Celoria 16, I-20133 Milano, Italy
}
\affiliation[b]{CERN, Theoretical Physics Department, CH-1211 Geneva 23, Switzerland}
\affiliation[c]{University of Jyvaskyla, Department of Physics, P.O. Box 35, FI-40014 University of Jyvaskyla, Finland}
\affiliation[d]{Helsinki Institute of Physics, P.O. Box 64, FI-00014 University of Helsinki, Finland}
\affiliation[e]{Department of Physics and Astronomy, Vrije Universiteit, NL-1081 HV Amsterdam}
\affiliation[f]{Nikhef Theory Group, Science Park 105, 1098 XG Amsterdam, The Netherlands}
\affiliation[g]{
  The Higgs Centre for Theoretical Physics, University of Edinburgh,\\
  JCMB, KB, Mayfield Rd, Edinburgh EH9 3JZ, Scotland
}

\preprint{CERN-TH-2024-015, Edinburgh 2024/4, Nikhef-2024-002}

\abstract{
  We present \yadism, a library for the evaluation of both polarized and unpolarized
  deep-inelastic scattering (DIS) structure functions and cross sections up to \nnnlo in
  perturbative QCD. The package provides computations of observables in fixed-flavor and
  zero-mass variable flavor number schemes. The implementation of the general
  mass variable flavor number schemes is supported through the high virtuality limits
  for the heavy flavor coefficients.
  In addition, \yadism provides a set of tools for the
  generation of interpolation grids in the PDF-independent \pineappl format, allowing to test
  the PDF dependence on any DIS observable without needing to rerun the computation.
  This work is part of an ongoing effort to standardize the format of theory predictions
  in high-energy physics within the \pineline framework. The code is open source, written
  in \texttt{Python} and documented to facilitate usage, integrations, and further extensions.
  Finally, the code has been benchmarked against the widely used \apfelpp and \qcdnum libraries.
}

\begin{document}
\maketitle
\flushbottom

\section{Introduction}
\label{sec:intro}
Deep-inelastic scattering (DIS) experiments provide strong constraints on the structure of hadrons, and around half of the experimental data points, used in the most recent PDF determinations
~\cite{Hou:2019efy,Bailey:2020ooq,NNPDF:2021njg}, corresponds to charged current (CC) or neutral current (NC) DIS
processes. These include both relatively recent data, such as the ones collected at HERA~\cite{H1:2009pze,H1:2018flt}, and earlier data such as from the BCDMS~\cite{Benvenuti:1989rh,Benvenuti:1989fm} experiment.
While in recent years the focus of particle physics phenomenology has shifted away from DIS in favor of the description of LHC data, the upcoming
Electron-Ion Collider (EIC) projects in the US~\cite{AbdulKhalek:2021gbh} and China~\cite{Anderle:2021wcy} have renewed interest in DIS.
Thus, an accurate description of these processes is required to optimally utilize their future data.
Reliable predictions for DIS are furthermore relevant for the interpretation of neutrino scattering data~\cite{Candido:2023utz}
from neutrino telescopes such as IceCube~\cite{IceCube:2014stg} and experiments such as the FPF~\cite{Anchordoqui:2021ghd,Feng:2022inv},
the LHeC~\cite{AbdulKhalek:2019mps,LHeC:2020van,LHeCStudyGroup:2012zhm} or FASER$\nu$~\cite{FASER:2019dxq} and
SND{@}LHC~\cite{SHiP:2020sos,SNDLHC:2022ihg} at the LHC.

In this paper we present \yadism, a new software library developed for the calculation of DIS observables with the requirements of the particle physics community
of the current age in mind. \Yadism differs from other \qcd codes such as \apfel~\cite{Bertone:2013vaa},
\apfelpp~\cite{Bertone:2017gds}, \hoppet~\cite{Salam:2008qg}, and \qcdnum~\cite{Botje:2010ay}
in various ways, which we briefly highlight in the following.

\Yadism includes most of the currently available results in literature, specifically it allows for the
computation of polarized~\cite{Hekhorn:2024tqm} and unpolarized
structure functions up to next-to-next-to-next-to-leading order (\nnnlo)~\cite{NNPDF:2024nan} in \qcd.
Thanks to its modular design, the library can be easily extended as the results of new computations become
available. The coefficients, whenever possible, have been benchmarked against \apfelpp and \qcdnum.

\Yadism provides both renormalization and factorization scale variations consistently~\cite{NNPDF:2024dpb} and both can be implemented at any order.
The currently implemented coefficients allow to perform renormalization scale variations up to \nnnlo and
factorization scale variations up to \nnlo. Instead, \nnnlo factorization scale variations can be included through the \eko
evolution code~\cite{Candido:2022tld}.

\Yadism can, together with \eko, be used to construct general-mass variable flavor number schemes (\gmvfns) using coexisting PDFs with different
numbers of active flavors. This can avoid~\cite{nFONLL} the perturbative expansion of the evolution kernel as is
currently done in the construction of the FONLL scheme~\cite{Forte:2010ta}.

\Yadism has a uniform treatment of all heavy quarks, i.e., all features that are available for charm are
also available for bottom and top.
This strategy opens up the possibility for computations with an
intrinsic bottom quark~\cite{Brodsky:2015fna,Lima:2024ilk}. We provide both the fixed-flavor number scheme (\ffns) and zero-mass variable-flavor number scheme (\zmvfns)
calculation, as well as the asymptotic limit, $Q^2 \gg m^2$, of the \ffns (\ffnz), which is required
in the construction of the FONLL scheme~\cite{Forte:2010ta}.

\Yadism is interfaced to
\pineappl~\cite{Carrazza:2020gss,christopher_schwan_2023_8422248}, a library
providing fast interpolation grids in a unique format and exposing an
Application Programming Interface (API) for the programming languages
\texttt{C}, \texttt{C++}, \texttt{Fortran}, \texttt{Rust}, and
\texttt{Python} making it portable and easy to use.
Fast interpolation grids provide a representation of predictions independent of
PDFs and the strong coupling, and therefore do not require rerunning the entire
toolchain of theory codes if one whishes to assess the impact of the \pdf on a
theory prediction. This feature can be useful both for \pdf fitting and also for
the determination of standard model parameters~\cite{Cridge:2023ztj}. Fast
interpolation grids were pioneered by FastNLO~\cite{Kluge:2006xs} and are a
vital part in the toolchain for phenomenology of perturbative QCD, as such the
grid technique has been adopted in various programs including
APPLgrid~\cite{Carli:2010rw} and APPLfast~\cite{Britzger:2022lbf}.
The \pineappl grid output format
allows \yadism to be integrated into the \texttt{xFitter}
framework~\cite{Alekhin:2014irh,xFitterDevelopersTeam:2017xal,xFitter:2022zjb}
and the \pineline framework~\cite{Barontini:2023vmr}. Specifically, the
latter consists of various codes with the aim to automate and efficiently
compute theory predictions for collider physics processes. Through this
toolchain, one can define a collection of consistent theory parameters and
observables of interest for which both the partonic coefficients along with
the DGLAP evolution kernels (through the \eko
package~\cite{Candido:2022tld}) can conveniently be calculated to produce
fast interpolation grids.

\Yadism is written in the \texttt{Python} programming
language, which is known for its ease of use, and thus reduces the threshold
for potential new contributors. For these reasons, development of new
functionality can be quick to, e.g., rapidly adopt new computations.
Proposed changes to the \yadism code are reviewed thoroughly and are
subjected to automated checks as part of a Continuous Integration (CI)
policy.

So far \yadism has already been used in various papers. Specifically, it has been used for the evaluation of neutrino structure functions in Refs.~\cite{Candido:2023utz,Cruz-Martinez:2023sdv}, and for the computation of polarized structure functions in Ref.~\cite{Hekhorn:2024tqm}.
Furthermore, \yadism has been adopted by the NNPDF collaboration, who has used it in their most recent papers~\cite{NNPDF:2024djq,NNPDF:2024dpb,NNPDF:2024nan}.

The \yadism code is open source and free to use under a GPL-3.0 license and can be found in its Github repository:
\begin{center}
  \url{https://github.com/NNPDF/yadism}
\end{center}
along with a user-friendly and up-to-date documentation:
\begin{center}
  \url{https://yadism.readthedocs.io/en/latest/}.
\end{center}
In this paper, we aim to provide an overview of some of the functionalities provided by \yadism while we refer the reader to the documentation for an
extensive overview of all the available features.

The rest of the paper proceeds as follows: in \cref{sec:theory}, we briefly summarize the theory underlying deep-inelastic scattering and provide
details on the \yadism implementation. In \cref{sec:bench}, we discuss representative benchmarks in comparison to other available libraries.
Finally, we conclude in \cref{sec:concl} and provide a brief description on possible extensions.
In addition, we include two appendices where we briefly comment on the calculation of a new set of formerly unknown coefficient functions in
\cref{sec:Kirill} and we give an explicit example on how to run \yadism in \cref{sec:user-manual}.

\section{The \yadism library}
\label{sec:theory}
In this section we introduce \yadism and provide an overview of its most important features.
In the following we assume the standard notation on DIS calculations as can
be found, e.g., in any textbook~\cite{pink-book} or in the PDG review~\cite{Workman:2022ynf}.

DIS structure functions can be evaluated within the framework of perturbative Quantum Chromodynamics (pQCD) using collinear factorization~\cite{Collins:1989gx} by
convoluting the PDFs with the relevant coefficient functions
\begin{equation}
  F\left(x, Q^2\right) = \sum_{j} \int_x^1 \frac{d z}{z} C_{j}\left(z, \alpha_s\left(Q^2\right)\right) f_j\left(\frac{x}{z}, Q^2\right),
  \label{eq:sf_in_qcd}
\end{equation}
where $j$ runs over all possible partons in the initial state, $f_j$ denotes the PDF of flavor $j$ and
$C_{j}$ are the coefficient function that can be calculated perturbatively as an expansion
in the strong coupling $\alpha_s$
\begin{equation}
  C_{j}\left(z, \alpha_s\left(Q^2\right)\right) = \sum_{k=0}^{\infty} \left( \frac{\alpha_s}{4 \pi} \right)^{k} C_{j}^{(k)}(z)\ .
  \label{eq:cf_in_qcd}
\end{equation}
Here and in the following we will assume that the \lo coefficient function are of $\mathcal{O}(\alpha_s^{0})$
irrespective of the first non-zero order.

All structure functions depend on two kinematic variables: the Bjorken-$x$ and the virtuality $Q^2$.
\cref{eq:sf_in_qcd} highlights how the DIS structure functions, describing the lepton-hadron interaction, depend on a linear combination of PDFs.
Thus, it is clear why DIS processes provide important constraints on flavor separation and are therefore fundamental for the PDFs determination
from experimental data.

However, while the factorization formula, \cref{eq:sf_in_qcd}, is conceptually simple, if one wishes to actually
compute a structure function one needs to define a number of theory parameters and parameters of the experimental setup.
Such input settings are passed to \yadism through \textit{runcards} in YAML format\footnote{\url{https://yaml.org/}}, and they are divided into two parts:
an \textit{observable runcard} describing the experimental setup (such as scattering particles, kinematic bins, or helicity settings)
and a \textit{theory runcard} describing the parameters of the theory framework (such as coupling strength, perturbative orders, or quark masses).
While observable runcards are usually tailored to a given experiment, theory parameters are usually shared by multiple runs.
An example of such runcards is reported in \cref{sec:user-manual}, where we show explicitly how to
run \yadism.

For completeness, the current settings of DIS datasets used in the
NNPDF framework are collected in the repository \textit{pinecards}:
\begin{center}
  \url{https://github.com/NNPDF/pinecards}\,,
\end{center}
where we specify the setup for a number of measurements at
NMC~\cite{NewMuon:1996uwk,NewMuon:1996fwh},
SLAC~\cite{Whitlow:1990gk,Whitlow:1991uw},
BCDMS~\cite{Benvenuti:1989rh,Benvenuti:1989fm},
CHORUS~\cite{CHORUS:2005cpn},
NuTeV~\cite{Goncharov:2001qe},
EMC~\cite{Aubert:1982tt},
and HERA~\cite{H1:2009pze,H1:2018flt}.
There, we also provide settings for several pseudo-measurements,
which are used as theoretical constraints in NNPDF~\cite{NNPDF:2021njg}.

Below, we describe the most important options for the configuration of the observables (\cref{sec:obs_options})
and theories (\cref{sec:th_options}) that can be defined in the respective runcard.
\cref{sec:tech_details} overviews the partonic coefficient functions implementation
and technical details on the computation thereof.

\subsection{Observable configuration options}
\label{sec:obs_options}

\paragraph{Projectile.}
\Yadism supports computations of DIS coefficients with massless charged leptons and their associated neutrinos as projectiles in the scattering process.
Specifically, to describe, e.g., the HERA data one needs both electrons and positrons
and, e.g., for the CHORUS data both neutrinos and anti-neutrinos are needed.
Charged leptons can interact both electromagnetically and weakly with the scattered nuclei, whereas neutrinos only
carry weak charges.
Recently, together with a machine-learning parametrization of experimental data,
CC neutrino DIS predictions computed with \yadism have been used to extend
predictions for neutrino structure functions~\cite{Candido:2023utz}.

\paragraph{Target.}
\Yadism supports computations with nuclei with mass number $A$ and $Z$ protons as targets in the scattering process.
By acting on the coefficients associated to up and down partons \yadism implements the isospin
symmetry of the form:

\begin{equation}
  \begin{pmatrix} c'_u \\ c'_d \end{pmatrix} =
  \frac{1}{A}
  \begin{pmatrix} Z & A - Z \\ A - Z & Z \end{pmatrix}
  \begin{pmatrix} c_u \\ c_d \end{pmatrix}
  \label{eq:isospin}
\end{equation}
where $c'_i$ and $c_i$ are the effective and the proton coefficient associated with the parton $i$.
This rotation is particularly useful in the context of proton PDF fitting
where it can be used to relate neutron, deuteron, and heavier nuclear structure functions to the
proton ones. In this way, isospin is used as a first approximation of nuclear correction by
just swapping up and down contribution for the amount specified by the target nuclei.
In particular for:
\begin{description}
  \item[proton targets]\textit{($A=1, Z=1$)}: up and down are kept as they are.
  \item[neutron targets]\textit{($A=1, Z=0$)}: up and down components are fully swapped,
  such that the up coefficient function is matched to the down PDF and conversely.
  \item[isoscalar targets, i.e. deuteron]\textit{($A=2, Z=1$)}:
  the effective coefficient functions will be mixed such that
  $c'_{u}$ will be half the original $c_{u}$ and half the original $c_{d}$.
\end{description}
\Yadism is completely general with respect to the nuclear target allowing a user to provide values for $A$ and $Z$ as input to the computation. Alternatively, for a number of targets, the name itself can also be provided as input.
The readily available targets are: iron ($A=49.618,Z=23.403$), used to describe NuTeV data;
lead ($A=208,Z=82$), used to describe CHORUS data; neon and marble ($CaCO_{3}$) with both $A=20, Z=10$,
used to describe respectively the BEBCWA59~\cite{BEBCWA59:1987rcd} and CHARM~\cite{CHARM:1984ikt} data.

\paragraph{Exchanged electroweak gauge boson(s).}
DIS can be categorized into three different processes defined by the gauge boson mediating the interaction:
\begin{description}
  \item[Electromagnetic current (EM)] corresponds to a process where the exchanged boson is a photon.
  \item[Neutral current (NC)] corresponds to cases where the exchanged boson does not carry any electric charge.
  Thus, it is a superset of the EM process where also the exchange of the $Z$ boson is allowed.
  Since for NC two bosons are allowed, interference terms must be included.
  Because the $Z$ boson has an axial coupling to the incoming lepton it introduces further complications related to $\gamma_5$~\cite{Gnendiger:2017pys}.
  Note that at small virtualities the $Z$ contributions are suppressed by a factor $Q^2/M_Z^2$.
  \item[Charged current (CC)] corresponds to processes where a $W^\pm$ boson is exchanged.
  The CC process is a flavor-changing current where the CKM-matrix encodes the probabilities to transition between different quark flavors~\cite{Cabibbo:1963yz,Kobayashi:1973fv}.
\end{description}

\paragraph{Cross sections.}
\Yadism supports the computation of both structure functions and (reduced) cross sections.
In particular, for the unpolarized scattering, we implement the structure functions:
\begin{equation}
  F_2, \ F_L, \ x F_3,
  \label{eq:unpol_sf}
\end{equation}
and their polarized counter parts:
\begin{equation}
  g_4, \ g_L, \ 2 x g_1,
  \label{eq:pol_sf}
\end{equation}
where the normalization is chosen such that at \lo,
all the structure functions are proportional to different \pdf combinations of the form $xf(x)$.

While structure functions may only depend on two variables,
cross sections may also depend on the inelasticity $y$.
Generally, we can write the (reduced) cross sections for a DIS process in terms of the structure functions as
\begin{equation}
  \sigma(x,Q^2,y)=N\left(F_2(x,Q^2)-d_L F_L(x,Q^2)+d_3 x F_3(x,Q^2)\right),
\end{equation}
where $N$, $d_L$, and $d_3$ may depend on the experimental setup or the scattered lepton. The different reduced cross sections implemented in \yadism, and their definitions in terms of $N$, $d_L$, and $d_3$ can be found in the online documentation\footnote{\url{https://yadism.readthedocs.io/en/latest/theory/intro.html\#cross-sections}}. The implemented definitions can be used
to describe data from
HERA, CHORUS, NuTeV, CDHSW~\cite{Berge:1989hr}, and FPF~\cite{Cruz-Martinez:2023sdv}.

Finally, we provide the linearly dependent structure functions:
\begin{equation}
  2x F_1 = F_2 - F_L, \quad \quad 2x g_5 = g_4 - g_L\,.
  \label{eq:f1_g1_def}
\end{equation}

\paragraph{Flavor tagging.}
In general, any total DIS structure function $F$ can be decomposed in three different components, according to the type of
quark coupling to the exchanged EW boson:

\begin{equation}
  F = F^{(l)} + F^{(h)} + F^{(hl)},
  \label{eq:sf_decomposition}
\end{equation}
where $F^{(l)}$ denotes the contribution coming from diagrams where all the fermion lines are massless,
$F^{(h)}$ is the contribution due to heavy quarks coupling to the EW boson and
$F^{(hl)}$ originates from higher order diagrams where a light quark is coupling to the boson, but
heavy quarks lines are present.

Given \cref{eq:sf_decomposition}, we support the calculation of fully inclusive (total) observables,
where only the lepton is observed in the final state, and flavor tagged final state,
where we require a specific heavy quark (charm, bottom, or top) to couple with the mediating boson.
This definition coincides with $F^{(h)}$ and it is an infrared-safe definition~\cite{Forte:2010ta}.
For example, the charm structure function $F^{(c)}$ can be obtained by assuming the coupling of any quark other than
charm and anti-charm to be zero.
Instead, a naive definition of $F^{(c)}$ by the heavy final state tag would not be infrared safe.
For completeness, also light structure functions $F^{(l)}$ are available,
in isolation, although they do not corresponds to any physical observable.

\subsection{Theory configuration options}
\label{sec:th_options}

\paragraph{Flavor number schemes.}
Flavor number schemes provide a prescription to resolve the ambiguous treatment of heavy quark masses.
Generally, to achieve a faithful description of experimental data
at scales roughly around the heavy quarks mass $Q\sim m$, quarks should be treated fully massive.
However, in the region where $Q \gg m$, quarks should be considered massless.
In \yadism we allow for 3 different schemes. Only one single heavy quark is allowed at each time.

\begin{description}
  \item[Fixed flavor number scheme (\ffns).]
  The \ffns, is defined as a configuration with a fixed number of flavors at all scales,
  i.e.\ all quark masses are fixed to be either light, heavy or decoupled.
  The \ffns retains all power-like heavy quark corrections $m^2/Q^2$ and a finite number
  of logarithmic corrections $\ln(Q^2/m^2)$.
  This finite number of logarithms, as opposed to a full resummation, limits the perturbative stable
  region.

  \item[Zero mass-variable flavor number scheme (\zmvfns).]
  In the \zmvfns all quark masses in the calculations are either light or decoupled.
  The number of light quarks $n_f$ is not fixed, but instead varies with the number of active flavors depending on the scale of the process, i.e.\ $n_f(Q^2)$. Specifically, $n_f=3$ below $m_c$ and this increases as the heavy quark thresholds are crossed, i.e.\ $Q>m_h$, after which the corresponding heavy quark is treated to be light.
  The \zmvfns resums all logarithmic corrections as they are provided by DGLAP evolution.
  However, the \zmvfns does not contain any power-like heavy quark corrections $m^2/Q^2$ which
  may be phenomenological important in certain regions of the kinematic phase space.

  \item[Asymtotic fixed flavor number scheme (\ffnz).]
  The \ffnz is similar to the \ffns, but retains only the logarithmic corrections,
  i.e.\ it does not contain any power-like heavy quark corrections $m^2/Q^2$.
  The \ffnz is constructed as the overlap between \ffns and \zmvfns and can be used to
  construct a \gmvfns flavor number schemes.

  A \gmvfns
  can be constructed to overcome the limitation due to potentially large missing corrections of \ffns and \zmvfns.
  One possible scheme is the FONLL scheme~\cite{Forte:2010ta}, which is defined through a linear combination of the \ffns and \zmvfns
  while taking care of possible double counting through the \ffnz.
  A detailed discussion on how to construct the FONLL scheme is given in Ref.~\cite{nFONLL}.
  \Yadism does not provide it explicitly, but all the necessary ingredients
  \ffns, \ffnz, and \zmvfns are available.

\end{description}

\paragraph{Renormalization and factorization scale variations.}
In perturbative QCD the coefficients $C_j$ of \cref{eq:sf_in_qcd},
are expanded in powers of $\alpha_s$.
The estimate of the error introduced by the truncation of such series
is quite relevant in multiple precision applications.
Some information about the missing higher orders, and the related uncertainty (\mhou),
can be extracted from the Callan-Symanzyk equations violation.
In this sense, a practical approach to obtain a numerical estimate consists in varying the relevant
unphysical scales.

In DIS, the two involved unphysical scales are the \textit{renormalization scale}, arising from the
subtraction of ultraviolet divergences, and the \textit{factorization scale}, from the
subtraction of collinear logarithms in the \pdf definition.

The explicit expressions of the $C_i$ expansion upon scale variations can be found, e.g., in Sec.\ 2 of Ref.~\cite{nnlo-sv-singlet}.
Generally, these depend, order by order in perturbation theory, on the derivatives
of $\alpha_s$ and the \pdf{}s with respect to the mentioned scales.
The former are the $\beta$-function coefficients and the latter the splitting functions.
In \yadism, necessary $\beta$-function coefficients are taken from the \eko package, while
the $x$-space splitting functions are directly implemented.

At the level of structure function, scale variations can be cast into an additional convolution with a kernel $K$:
\begin{equation}
  F(x,\mu\neq Q) = \left(K \otimes C \otimes f\right)(x)
\end{equation}
It can be shown that the transformation can be applied a-posteriori to an already computed interpolation grid.

\paragraph{Target mass corrections.}
While \cref{eq:sf_in_qcd} is usually derived for the scattering of two massless particles,
it is possible to account for the finite mass of the scattering target through target mass corrections~\cite{tmc-review,tmc-iranian,Hekhorn:2024tqm}.
These corrections become relevant for either small virtualities or large Bjorken-$x$.
They can be implemented as an additional convolution and we provide several approximations (corresponding to higher twist expansions)
following Ref.~\cite{tmc-review}.

\subsection{Implemented partonic coefficients and computation details}
\label{sec:tech_details}

\paragraph{Quark mass corrections.}
We can differentiate quarks into three different types:
light ($m=0$), heavy ($m$ finite) and decoupled ($m=\infty$).
Thus, each coefficient function of~\cref{eq:sf_in_qcd} can be categorized by the appearance of heavy quark lines in various parts of the diagrams:
\begin{description}
  \item[Light] does not contain massive corrections in any part, i.e.\ all quarks are either light or decoupled.
  \item[Heavy] contains heavy quarks in the output. Note that while some calculation of coefficients with two mass scales are available~\cite{Ablinger:2018nby},
    in \yadism{} we currently only provide support for coefficients depending on a single heavy quark mass scale since the impact of the missing
    corrections are small.
  \item[Intrinsic] contains contributions where the incoming parton is a heavy quark and which thus allows for intrinsic heavy quarks as opposed to
    radiatively produced heavy quarks.
  \item[Asymptotic] is the $Q\gg m$ limit of either the heavy or intrinsic coefficient. The asymptotic contributions are used in the construction of general mass variable
    flavor number schemes.
\end{description}

This classification is not exclusive
and it is useful to only distinguish coefficient functions but
it does not correspond to a unique trivial mapping at the level of
structure functions (see \cref{eq:sf_decomposition}).
In fact, depending on the chosen variable flavor number scheme, the same
coefficients can be reshuffled differently inside each of the components $F^{(l)}$, $F^{(h)}$, and $F^{(hl)}$,
or might even not be present at all. For example in the FONLL scheme~\cite{Forte:2010ta}
$F^{(c)}$ is computed with massive quarks (using \textit{heavy}), with massless quarks (using \textit{light}),
and in the asymptotic mass limit (using \textit{asymptotic}).

\paragraph{Partonic coefficient functions.}

\begin{table}
  \centering
    \renewcommand{\arraystretch}{1.6}
    \begin{tabular}{c | c c c c}
    \toprule
    NLO $O(a_s)$ & light & heavy & intrinsic & asymptotic  \\
    \hline
    NC & \grokcell\cite{vogt-f2nc,vogt-flnc,moch-f3nc} & \grokcell\cite{Hekhorn:2019nlf} & \grokcell\cite{kretzer-schienbein} & \grokcell\cite{Buza:1996wv}\\
    CC & \grokcell\cite{vogt-f2lcc,vogt-f3cc} & \grokcell\cite{gluck-ccheavy} & \grokcell see \cref{sec:Kirill} & \grokcell\cite{Buza:1996wv}\\
    \midrule
    NNLO  $O(a_s^2)$  & & &\\
    \hline
    NC & \grokcell\cite{vogt-f2nc,vogt-flnc,moch-f3nc} & \grokcell\cite{Hekhorn:2019nlf} & \rdxcell & \grokcell\cite{Buza:1996wv}\\
    CC & \grokcell\cite{vogt-f2lcc,vogt-f3cc} & \ylcell \cite{Gao:2017kkx}\ftm{1} & \rdxcell & \ylcell\cite{Buza:1996wv}\\
    \midrule
    N$^3$LO $O(a_s^3)$ & & &\\
    \hline
    NC & \grokcell\cite{vogt-f2nc,vogt-flnc,moch-f3nc} &  \grokcell\cite{Kawamura:2012cr,Laurenti:2024anf,bbl2023} & \rdxcell & \grokcell\cite{Bierenbaum:2008yu,Bierenbaum:2009mv,Ablinger:2010ty,Ablinger:2014vwa,Ablinger:2014nga,Behring:2014eya,bbl2023} \\
    CC & \grokcell\cite{vogt-f2lcc,vogt-f3cc} &  \rdxcell & \rdxcell & \rdxcell \\
    \bottomrule
  \end{tabular}
  {
    \footnotesize
    \renewcommand{\arraystretch}{1.2}
    \begin{tabular}{r l}
      \fnsym{1} & Available as $K$-factors.
    \end{tabular}
  }
  \vspace{0.2cm}
  \caption{Overview of the unpolarized DIS coefficients currently implemented in the \yadism{} library at the corresponding order in perturbative QCD.
    In the columns we distinguish between light, heavy, intrinsic, and asymptotic.
    We mark in green coefficient function that are implemented in \yadism, in red the ones which are not yet known and
    in yellow the ones which are not yet implemented in \yadism, but available in literature.}
  \label{tab:coef-funcs}
\end{table}

\Yadism implements both unpolarized
and polarized coefficient functions up to \nnnlo in fixed-order QCD.
In \cref{tab:coef-funcs,tab:pol-coeff-funcs} we collect a summary of the coefficient functions
as currently implemented.
For each perturbative order\footnote{Recall that we adopt an \textit{absolute}
terminology of perturbative order, i.e.,
\lo $=O(a_s^0)$ irrespective of the first non-zero order, e.g.\ for
$F_L$ or $F_2^{(c)}$.}
and process, we distinguish contributions from light, heavy, asymptotic
and intrinsic coefficients.
In the unpolarized case, \cref{tab:coef-funcs}, the light, heavy, and asymptotic contributions are available up to \nnnlo,
except for the CC.
The intrinsic components are also available only up to
NLO, with the CC part computed very recently, see also \cref{sec:Kirill}.
The NNLO corrections to CC are known only through $K$-factors~\cite{Gao:2017kkx} and their implementation into \yadism
is currently work in progress.
Instead, for the heavy NC \nnnlo coefficients, a full analytical result is not known, although approximations
can be constructed using information already available through resummations and high virtuality limits~\cite{Kawamura:2012cr,Laurenti:2024anf}.
These coefficient are then implemented together with an uncertainty which can be propagated to the final result.

\begin{table}
  \centering
  \renewcommand{\arraystretch}{1.60}
  \begin{tabular}{c|c c c c}
    \toprule
    & light & heavy & intrinsic & asymptotic  \\
    \hline
    NLO  $O(a_s)$ & \grokcell\cite{Zijlstra:1993sh,deFlorian:1994wp,Anselmino:1996cd} & \grokcell\cite{Hekhorn:2019nlf} & \rdxcell & \grokcell\cite{Buza:1996xr} \\
    \midrule
    NNLO  $O(a_s^2)$ & \grokcell\cite{Zijlstra:1993sh,Borsa:2022irn} & \grokcell\cite{Hekhorn:2019nlf} & \rdxcell & \grokcell\cite{Buza:1996xr} \\
    \midrule
    N$^3$LO $O(a_s^3)$ & \ylcell\cite{Blumlein:2022gpp}\ftm{1} &  \rdxcell & \rdxcell & \ylcell\cite{Behring:2015zaa,Ablinger:2019etw,Behring:2021asx,Blumlein:2021xlc,Bierenbaum:2022biv,Ablinger:2023ahe}\ftm{1} \\
    \bottomrule
  \end{tabular}
  {
    \footnotesize
    \renewcommand{\arraystretch}{1.2}
    \begin{tabular}{r l}
      \fnsym{1} & Only for the $g_1$ structure function.\\
    \end{tabular}
  }
  \vspace{0.2cm}
  \caption{
    Same as \cref{tab:coef-funcs} for NC polarized coefficients.
  }
  \label{tab:pol-coeff-funcs}
\end{table}

Similarly, \cref{tab:pol-coeff-funcs} provides an overview of the polarized NC coefficient functions
currently implemented.
As opposed to the unpolarized counterpart, intrinsic polarized coefficient functions are not yet known.
At \nnnlo, only the light coefficient functions for the structure function $g_1$ have been computed~\cite{Blumlein:2022gpp} together with
the heavy asymptotic limit $Q^2 \gg m_h^2$~\cite{Behring:2015zaa,Ablinger:2019etw,Behring:2021asx,Blumlein:2021xlc,Bierenbaum:2022biv,Ablinger:2023ahe}.
Their implementation in \yadism is left for future updates.

\paragraph{Analytic structure of coefficient functions.}
The coefficient functions are not restricted to being regular functions, but they
might also correspond to a Dirac delta function or singular distributions.
In particular, the latter occur in the light coefficient functions because of massless quarks:
whenever the mass of the quark does not prevent IR divergences,
it generates \emph{plus distributions} upon subtractions.

The presence of such distributions does not cause any issue at the analytical level,
since the coefficients have to be convoluted with a \pdf (or an interpolation polynomial as in \cref{eq:FK}),
and thus they always act within the scope of an integral.
Instead, it does require a dedicated treatment at numerical level, since a distribution
cannot be just evaluated (sampled) at given points, and integrated with some
approximation, as it is done for regular functions.
As common in literature, our $x$-space coefficient function implementation
follows the so called \textit{regular, singular, local} formalism, first described in \cite{Floratos:1981hs}.

\paragraph{Grid formalism.}
It is common for \dis calculations to provide coefficient functions that are directly convoluted with a given PDF, thus returning the predicted value for the requested DIS cross section.
This is, however, not necessarily the most practical approach. Instead one may wish to store the computation of the DIS coefficient in an interpolation grid format, thus factorizing the \pdf dependence.
This is useful in situations where predictions for the same observable have to be computed for different PDFs. The case where this is clearly most relevant is in the context of \pdf fits, where at each step of the fitting procedure new comparisons to data are required.
In order to unify the treatment inside a \pdf fit
we follow the \pineline framework~\cite{Barontini:2023vmr} and provide
interpolation grids, which are more beneficial in a PDF fitting environment.

We introduce an interpolation for the PDF $f(x)$ using the nodes $x_k$ and
its associated basis of interpolation polynomials $p_k(x)$ and write
\begin{equation}
  f(x) = \sum_k f(x_k) p_k(x) = f^k p_k(x)
\end{equation}
where we defined $f^k \equiv f(x_k)$ and, as usual, sum over repeated indices.
While the choice of the nodes is left up to the user, the interpolation basis is fixed to
piece-wise Lagrange polynomials, and provided by \eko.
The \pdf values can then be evaluated directly on the nodes from the original
parametrization, or (re-)interpolated from distributed \pdf grids
(such as provided by \lhapdf \cite{Buckley:2014ana}).

To compute an observable $\sigma(x)$ for a given Bjorken-$x$, we can then write
\begin{align}
  \sigma(x) = (C \otimes f)(x) = f^k (C \otimes p_k)(x) = f^k C_k(x) \label{eq:FK}
\end{align}
and identify $C_k(x) = (C\otimes p_k)(x)$ as the sought-after interpolation grid.
Note that in \cref{eq:FK} we suppressed for the sake of readability the flavor dependency,
the scale dependency, and the dependency on the strong coupling.
In practice, however, we need to keep track of all of them and the \pineappl format~\cite{Carrazza:2020gss,christopher_schwan_2023_8422248}
supports such a full breakdown.

\section{Benchmarking and Validation}
\label{sec:bench}
Having described the \yadism library and its available features, we will now provide
various benchmark analyses. First, we benchmark \yadism with some of the most widely used libraries for the computation of DIS observables,
namely \apfelpp and \qcdnum.
Then we provide representative comparisons on the different prescriptions used to treat
heavy quark masses in order to underline their relevance in the different kinematic regions.
In all the subsequent comparisons, we adopt a fixed boundary condition defined as a PDF set at a given scale $Q=Q_0$. Evolution of the boundary condition, including changing of the number of active flavors, is performed using \eko.

\subsection{Benchmarking}

Let us start discussing the benchmarks of \yadism in comparison to other available libraries for a set of representative structure
functions using various flavor number schemes and different perturbative orders.
In particular, we show benchmarks for both the unpolarised structure function $F_{2}$
and its polarised counterpart $2xg_1$ using \zmvfns\footnote{While \zmvfns allows for a variable number of active flavors, i.e. $n_f(Q^2)$, here, and in the rest of this section, we keep $n_f$ fixed to simplify the discussion.} to highlight
the accuracy of the massless calculation. Then we compare $F_{2}^{\text{(c)}}$ with \ffns with $n_f=3$
light flavors to highlight heavy quark mass effects.
As benchmark tools, we rely on two main programs:
\begin{description}
  \item[\rm \apfelpp~\cite{Bertone:2017gds}] which provides \dis observables up to \nnnlo for massless, unpolarized structure functions and
    up to \nnlo for massless, polarized structure functions.
    It extends the functionalities of the previous Fortran code \apfel{}~\cite{Bertone:2013vaa}
    and has an explicit dependence on the \pdf, which can be interfaced via \lhapdf~\cite{Buckley:2014ana}.
  \item[\rm \qcdnum~\cite{Botje:2010ay}] which computes \dis structure functions up to \nnlo for
    unpolarized parton densities and up to \nlo for polarized parton densities.
    The program implements both \ffns and \zmvfns and uses
    polynomial spline interpolation to compute the structure function from a given \pdf.
\end{description}

The results reported below show the agreement between \yadism
and other tools.

\paragraph{Massless coefficient functions.}
\begin{figure}
  \centering
  \includegraphics[width=.48\textwidth]{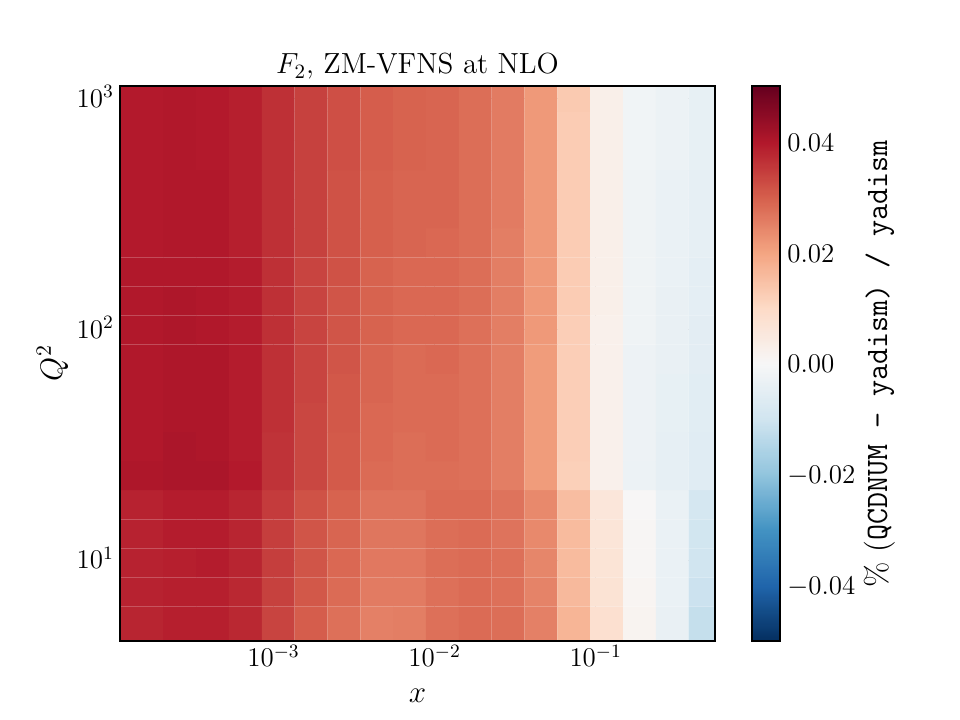}
  \includegraphics[width=.48\textwidth]{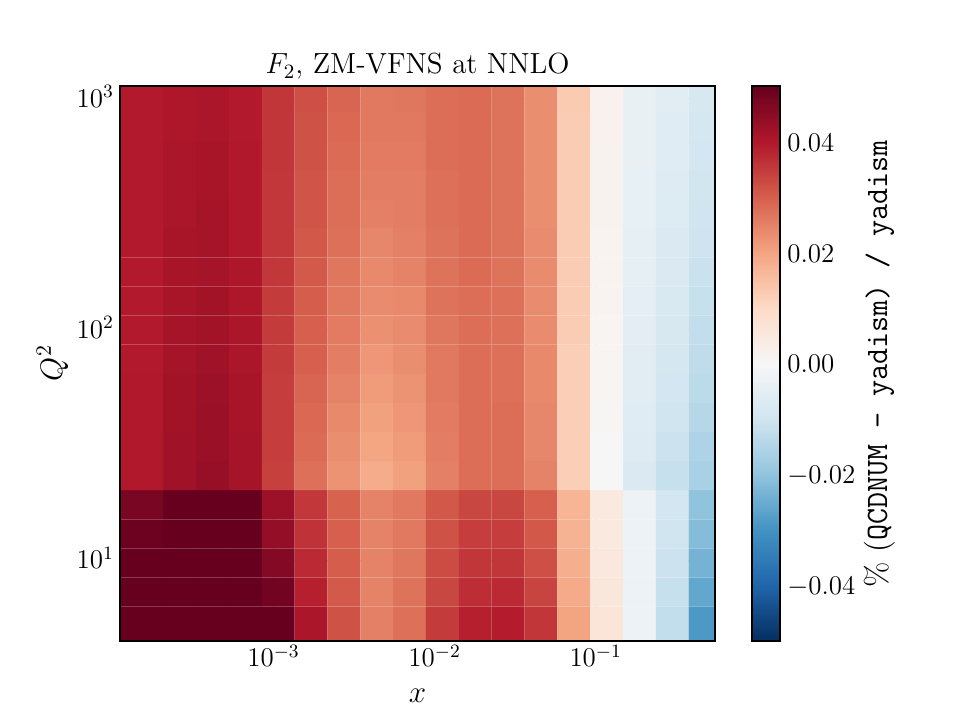}
  \caption{ Relative difference between \yadism and \qcdnum for the
  structure function $F_{2}$ using \zmvfns as function of $x$ and $Q^2$ at
  \nlo (left) and \nnlo (right) accuracy.
  }
  \label{fig:f2_light_qcdnum}
\end{figure}
\begin{figure}
  \centering
  \includegraphics[width=.48\textwidth]{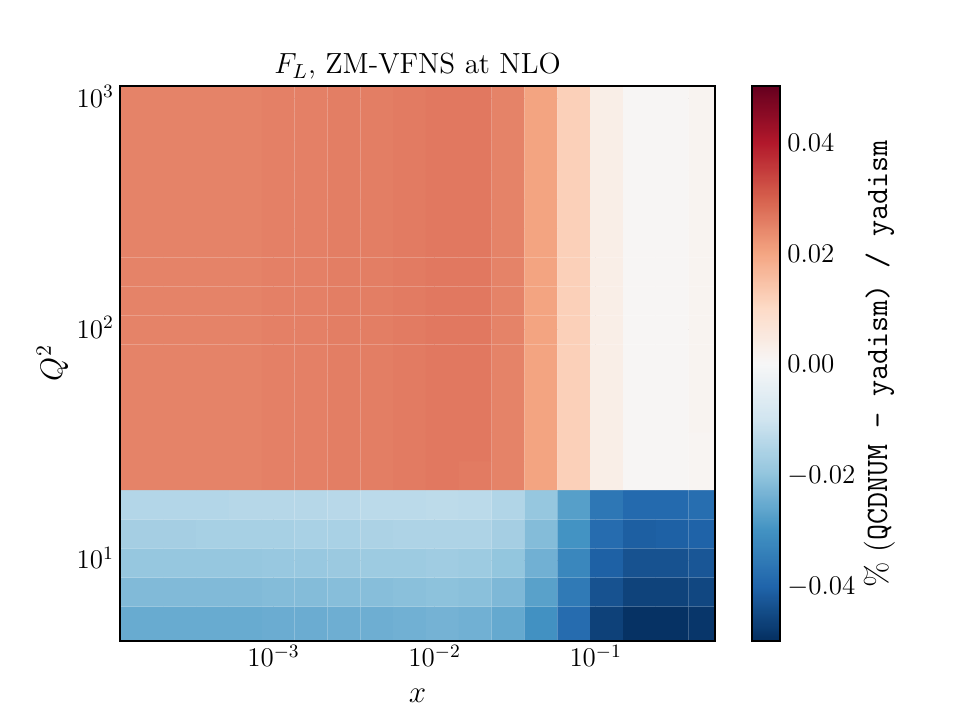}
  \includegraphics[width=.48\textwidth]{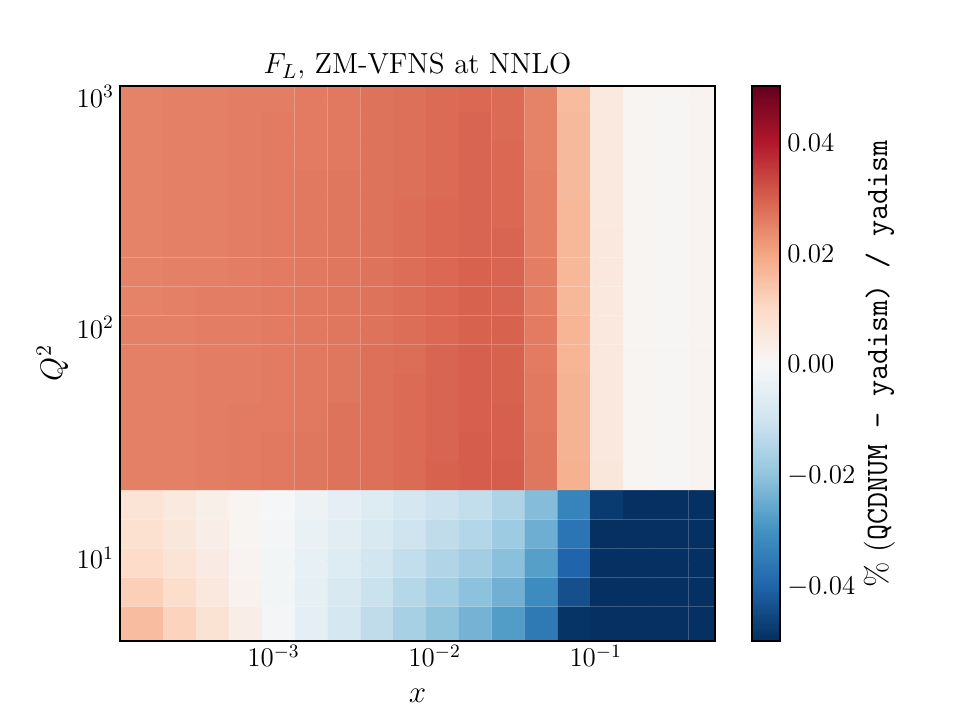}
  \caption{Same as \cref{fig:f2_light_qcdnum}, but now comparing the structure function $F_L$.
  }
  \label{fig:fl_light_qcdnum}
\end{figure}
In the \zmvfns only massless coefficient functions are involved, thus
we expect to reach good agreement with different tools for a broad range of kinematics.
Here we select $x \in [10^{-4}, 1]$ and $Q^2 \in [4.0, 10^4]\,\si{\GeV^2}$, covering the relevant ranges
for \dis phenomenology studies.
For simplicity, we focus on NC structure functions,
but analogous results hold also for CC \dis.
First, in \cref{fig:f2_light_qcdnum} (\cref{fig:fl_light_qcdnum}) we show the relative difference on $F_{2}$ ($F_L$)
between \yadism and \qcdnum computed at \nlo (left) and \nnlo (right) accuracy for
different kinematics ranges.
The overall agreement is within $\SI{0.05}{\percent}$ with the largest discrepancies
visible in the small-$x$ corner.

\begin{figure}
  \centering
  \includegraphics[width=.48\textwidth]{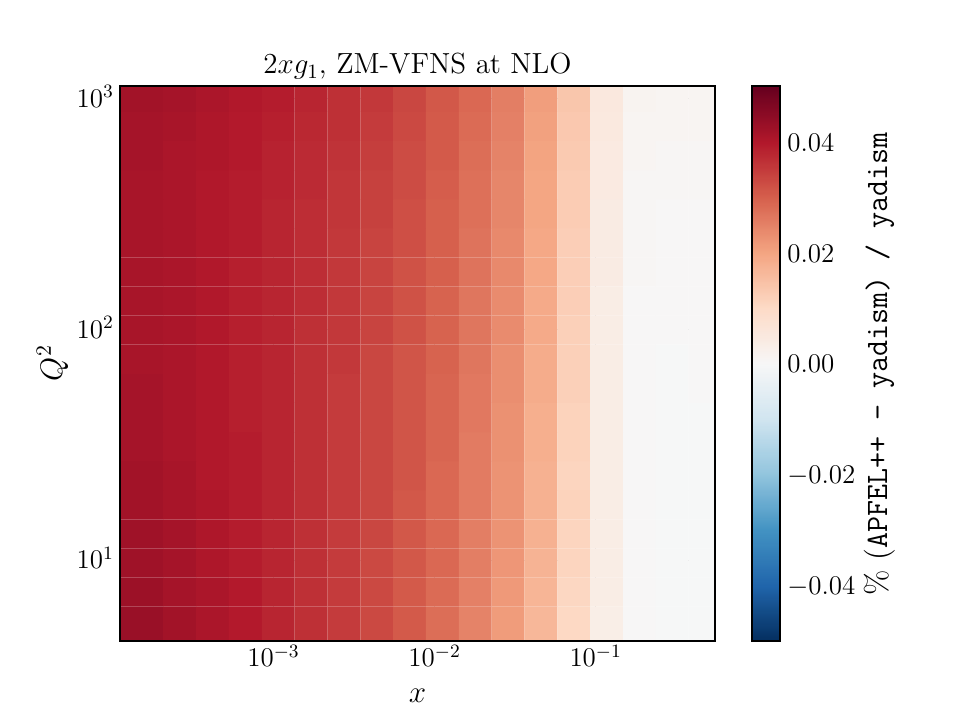}
  \includegraphics[width=.48\textwidth]{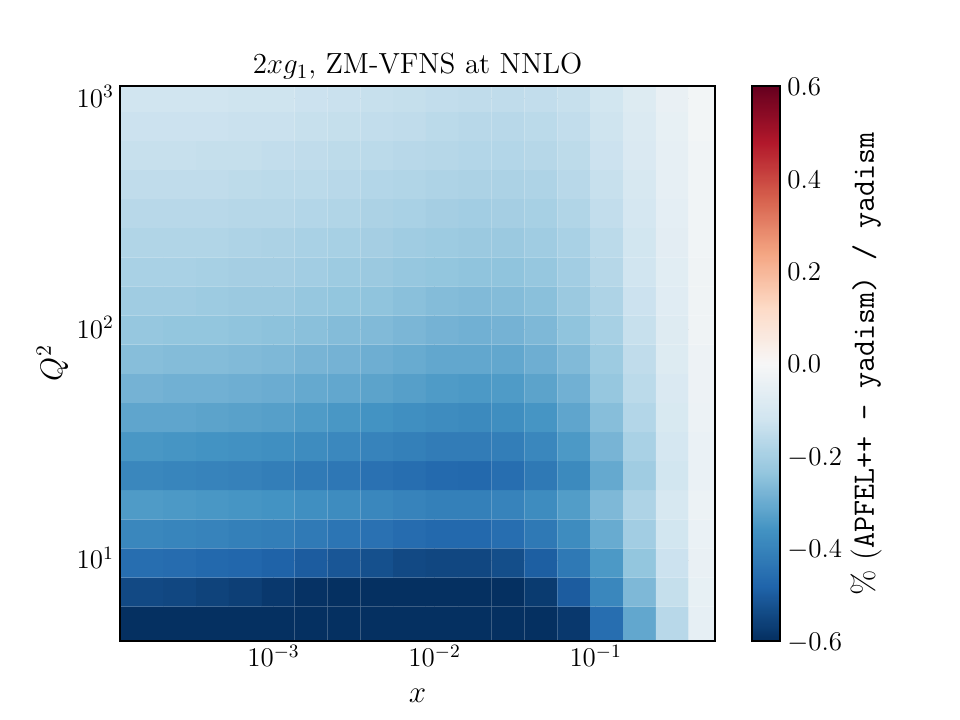}
  \caption{
    Same as \cref{fig:f2_light_qcdnum}, but now comparing the structure function
    $2 x g_1$ computed with \yadism and \apfelpp.
  }
  \label{fig:g1_total_apfelpy}
\end{figure}
\begin{figure}
  \centering
  \includegraphics[width=.48\textwidth]{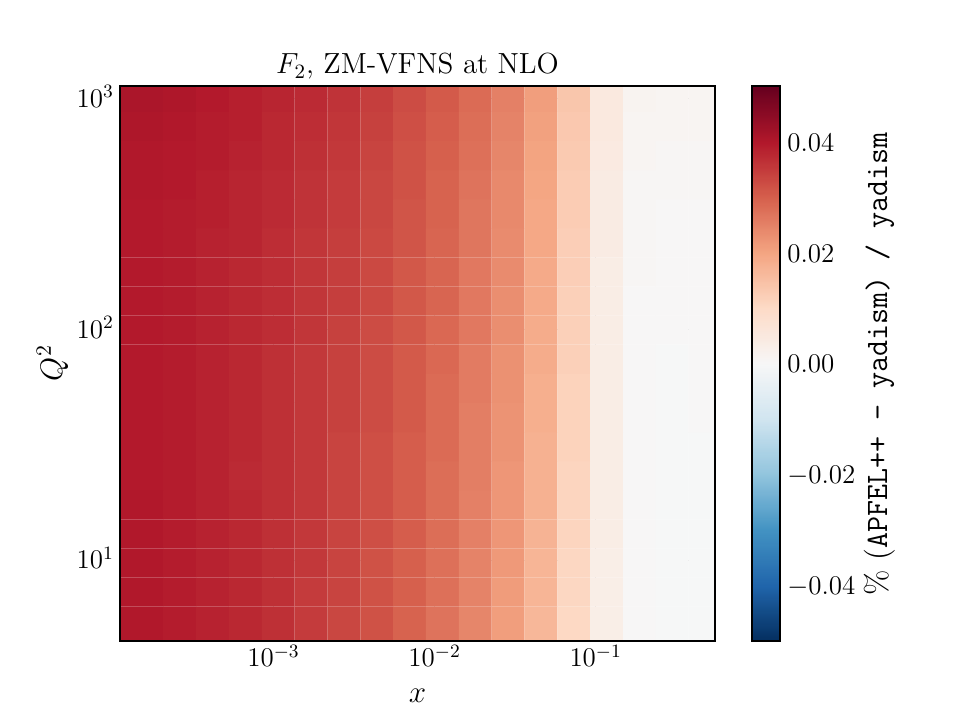}
  \includegraphics[width=.48\textwidth]{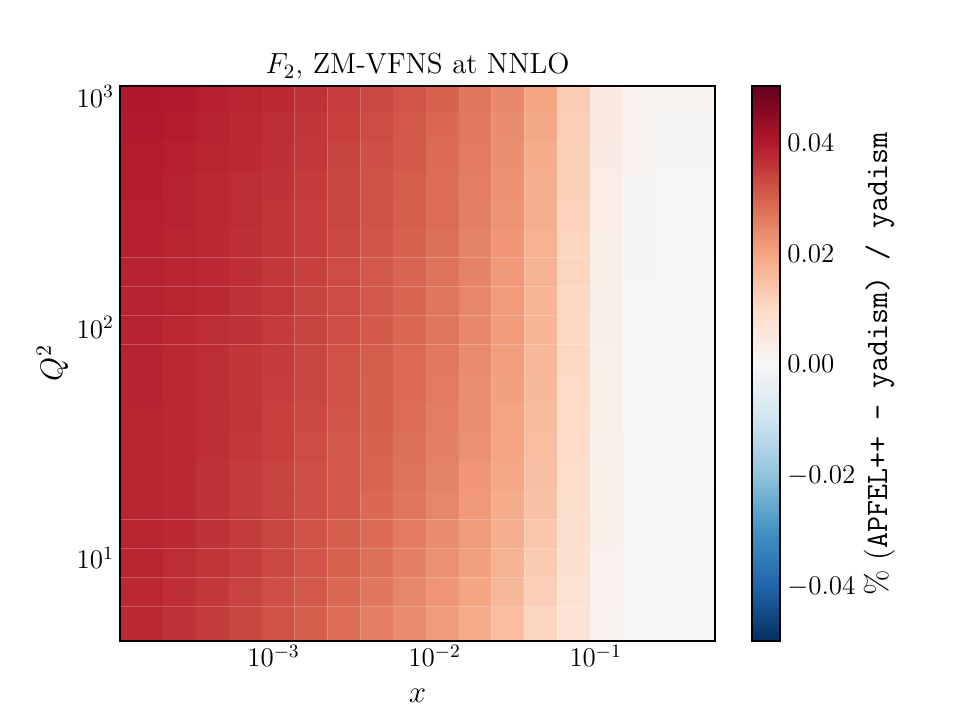}
  \includegraphics[width=.48\textwidth]{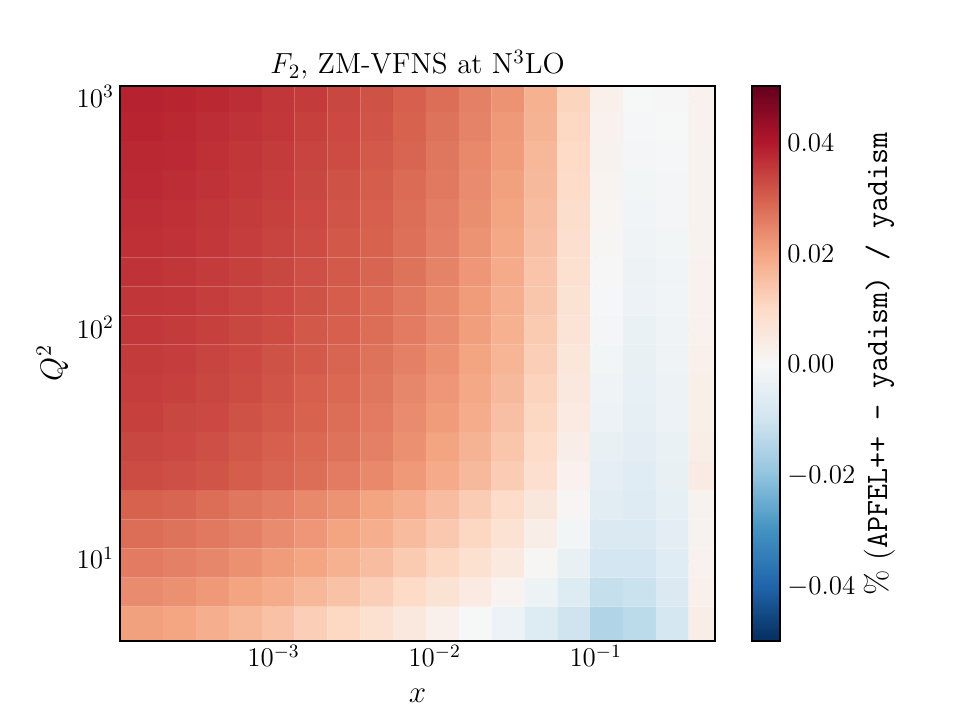}
  \caption{
    Same as \cref{fig:f2_light_qcdnum}, but now comparing the structure function
    $F_2$ computed with \yadism and \apfelpp at \nlo (upper left),
    \nnlo (upper right) and \nnnlo (bottom) accuracy.
  }
  \label{fig:f2_total_apfelpy}
\end{figure}
\begin{figure}
  \centering
  \includegraphics[width=.48\textwidth]{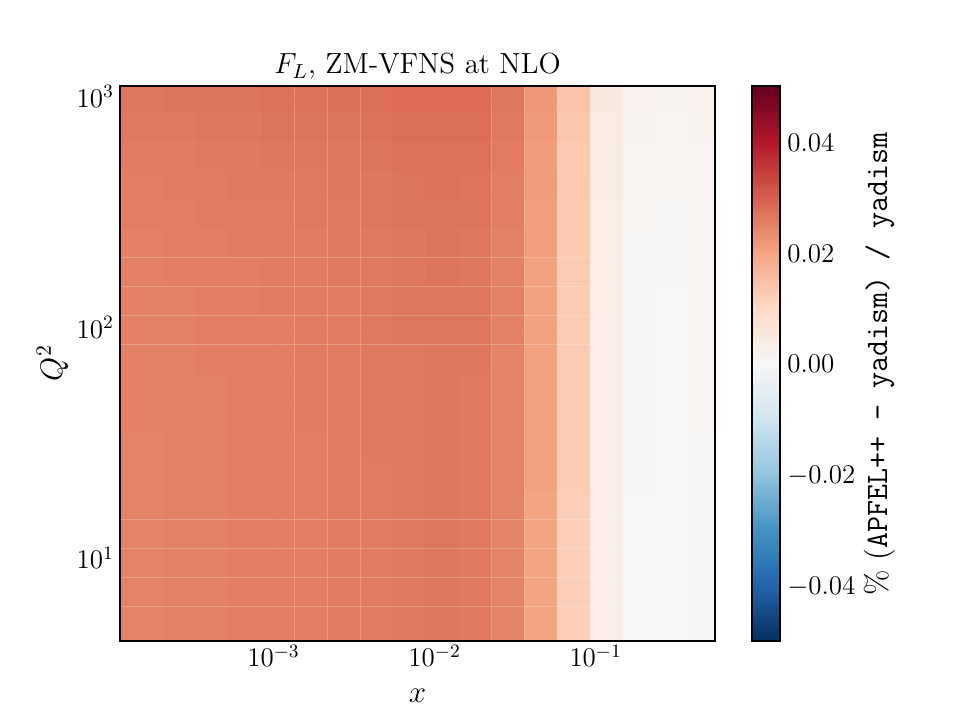}
  \includegraphics[width=.48\textwidth]{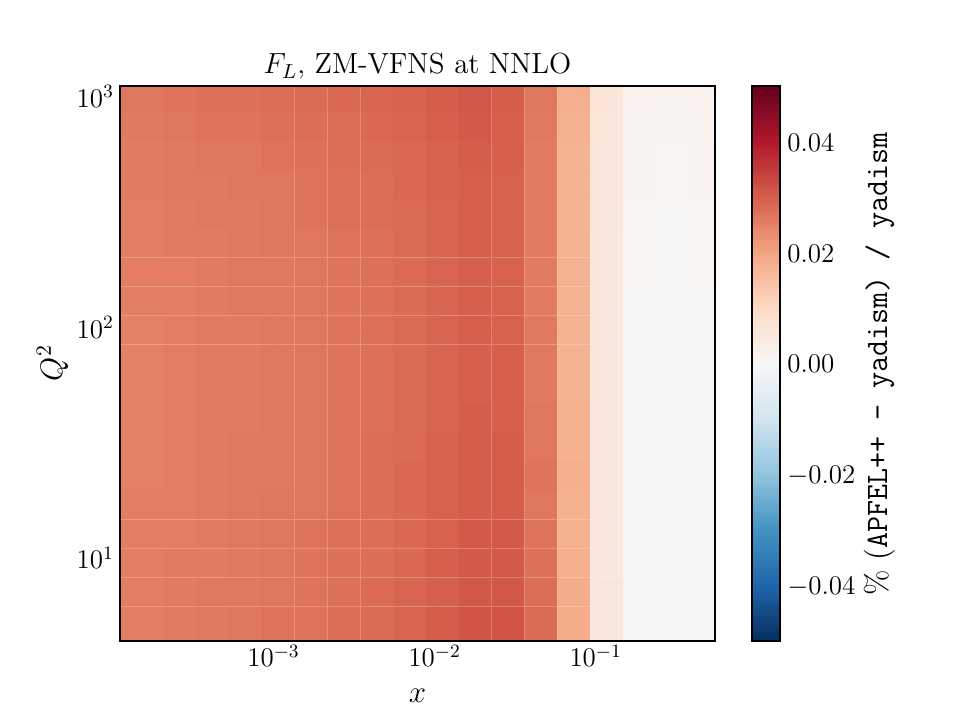}
  \includegraphics[width=.48\textwidth]{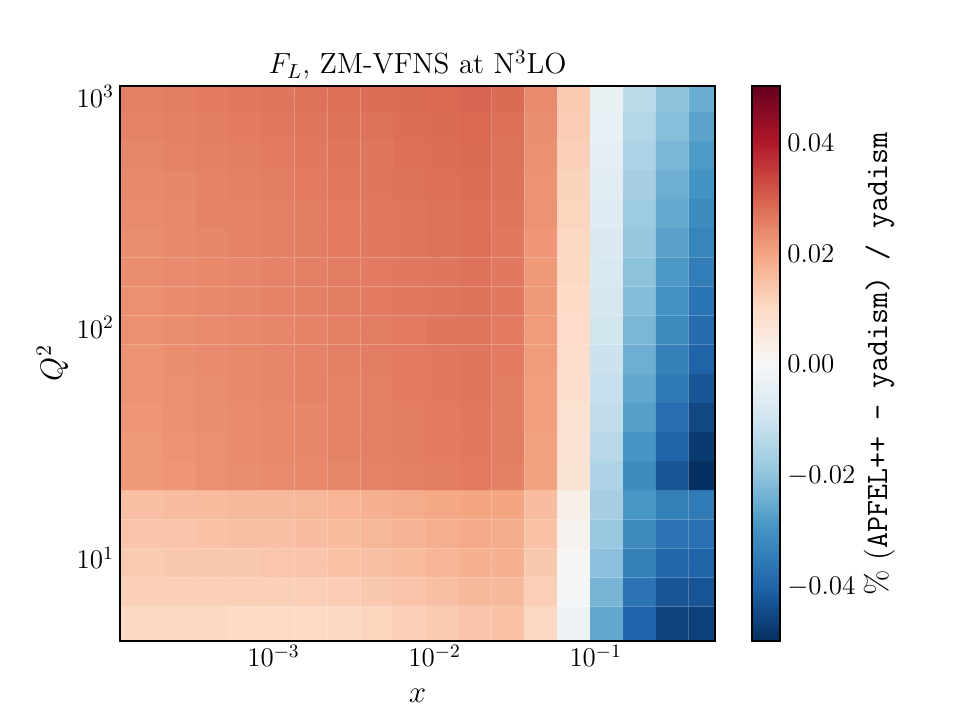}
  \caption{
    Same as \cref{fig:f2_total_apfelpy}, but now comparing the structure function $F_L$.
  }
  \label{fig:fl_total_apfelpy}
\end{figure}

The analogous comparison with \apfelpp is displayed in \cref{fig:g1_total_apfelpy}
for the polarized structure function $2 x g_1$
and in \cref{fig:f2_total_apfelpy} (\cref{fig:fl_total_apfelpy}) again for $F_{2}$ ($F_L$) but now
at \nlo (upper left), \nnlo (upper right) and \nnnlo (bottom) accuracy.
Also here the agreement between the different codes is always within $\SI{0.05}{\percent}$.
An exception is found for $2 x g_1$ at \nnlo,
where the differences are around $\SI{0.5}{\percent}$. This larger difference is a result of the different implementation of the nonsinglet (NS) coefficient function --
while \yadism exploits the exact symmetry of $\Delta C_{1,\mathrm{NS}}^{(2)} = C_{3,\mathrm{NS}}^{(2)}$
\cite{Borsa:2022irn}, \apfelpp implements the analytical calculation
from \cite{Zijlstra:1993sh}.

From the examples discussed, it is clear that the accuracy of the results does not depend on the perturbative order,
i.e.\ the pattern is not affected by the complexity of the calculation.

\paragraph{Heavy quark mass effects.}
Benchmarks of massive calculations are more involved because massive effects are subdominant in most of the kinematic regions,
and can be affected by different approximation of the massive coefficient
functions~\cite{Hekhorn:2019nlf,Beenakker:1988bq}.

In order to verify the accuracy of our implementation,
we report the comparison for the EM charm-tagged structure function $F^{\text{(c)}}$,
computed in \ffns with three light flavors.
We adopt the same kinematic range in $Q^2$ as in the previous part,
but we select $x \in [10^{-4}, 10^{-1}]$ excluding the large-$x$ region
where massive structure functions become small and relative uncertainties large.
Moreover, a sufficiently large-$x$ corresponds to an energy that is below the threshold to produce a heavy quark pair ($s < 4m^2$).

\begin{figure}
  \centering
  \includegraphics[width=.48\textwidth]{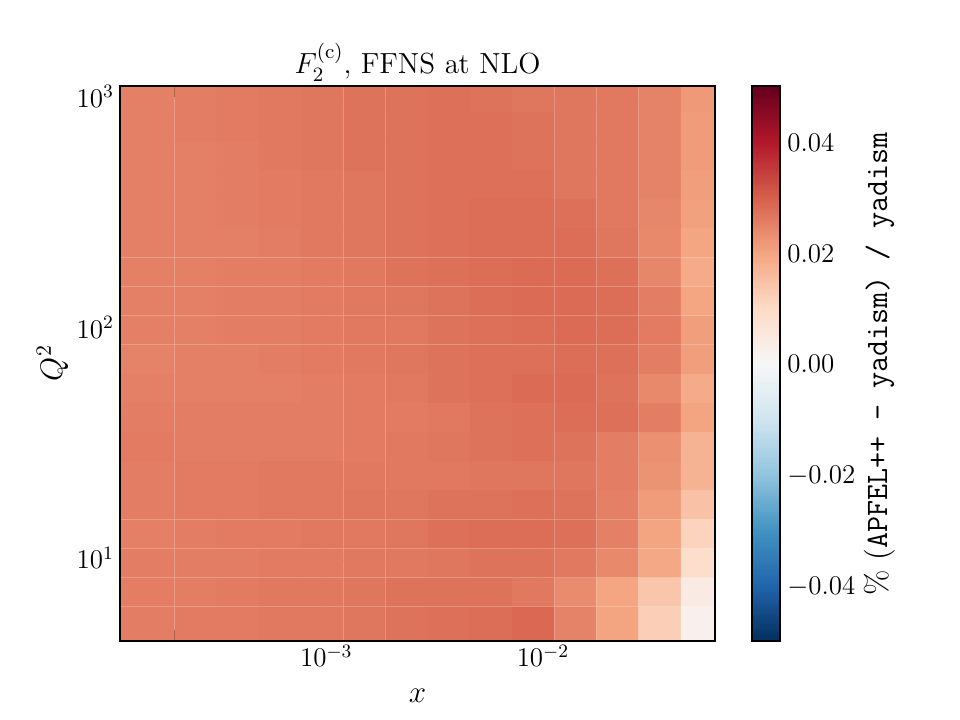}
  \includegraphics[width=.48\textwidth]{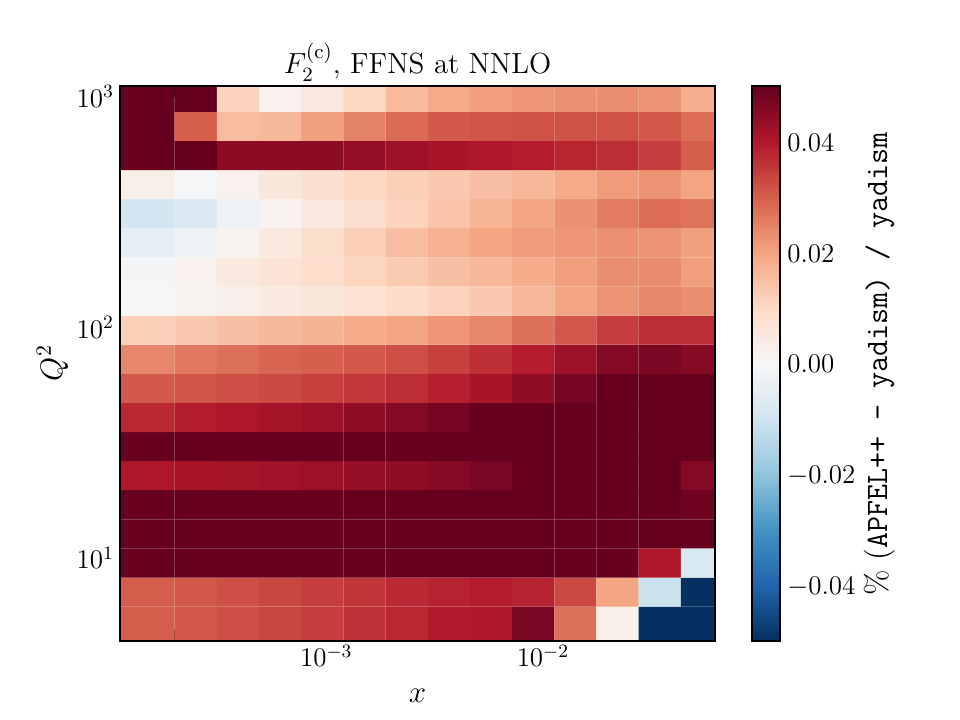}
  \caption{ Relative difference between \yadism and \apfelpp for the
  structure function $F_{2}^{\text{(c)}}$ using \ffns, $n_f=3$, as function of $x$ and $Q^2$ at
  \nlo (left) and \nnlo (right) accuracy.
  }
  \label{fig:f2_charm_apfelpy}
\end{figure}

\begin{figure}
  \centering
  \includegraphics[width=.48\textwidth]{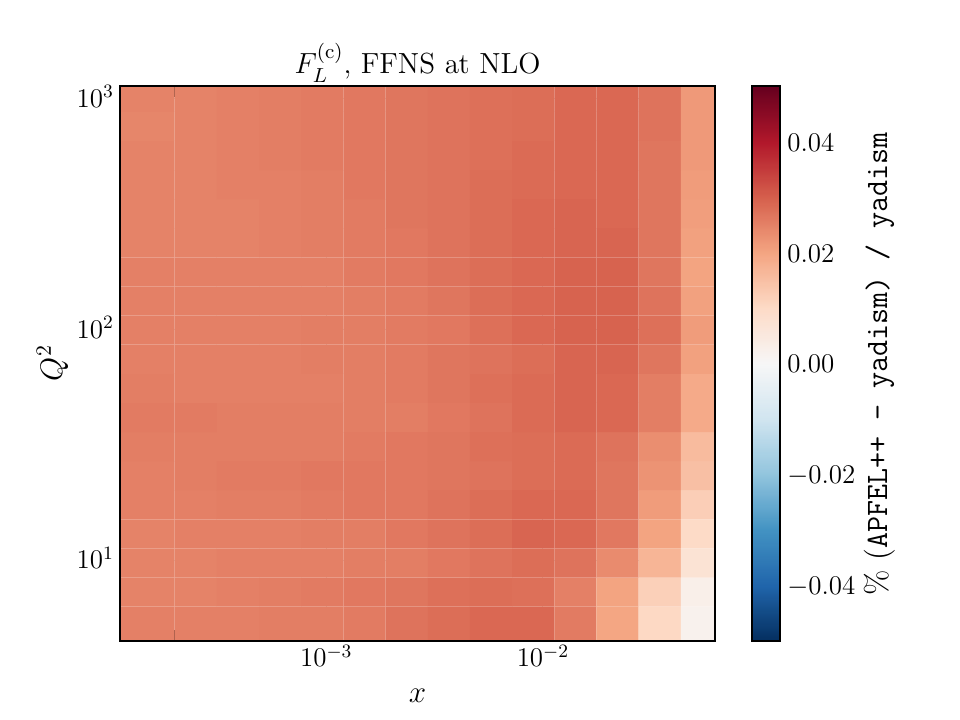}
  \includegraphics[width=.48\textwidth]{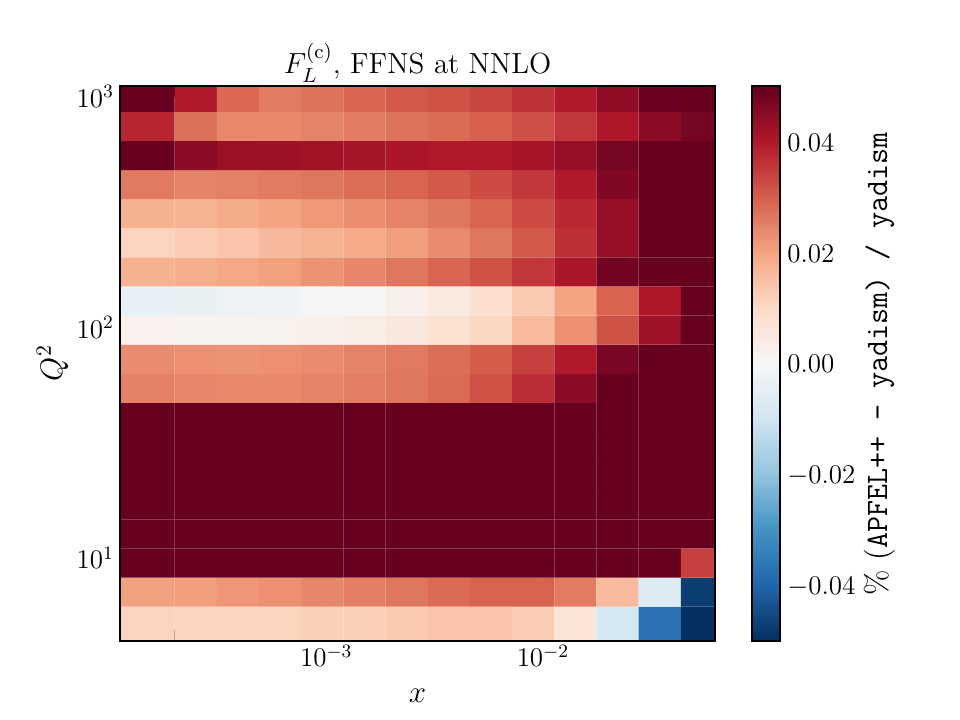}
  \caption{
    Same as \cref{fig:f2_charm_apfelpy}, but now comparing the structure function $F_L^{(c)}$.
  }
  \label{fig:fl_charm_apfelpy}
\end{figure}

\cref{fig:f2_charm_apfelpy} (\cref{fig:fl_charm_apfelpy}) displays the relative difference between \apfelpp
and \yadism for a \nlo and \nnlo computations of $F_2^{\text{(c)}}$ ($F_L^{\text{(c)}}$).
In this case, the agreement is around $\SI{0.02}{\percent}$ at \nlo
for most of the kinematics and around $\SI{0.05}{\percent}$ at \nnlo.

\begin{figure}
  \centering
  \includegraphics[width=.48\textwidth]{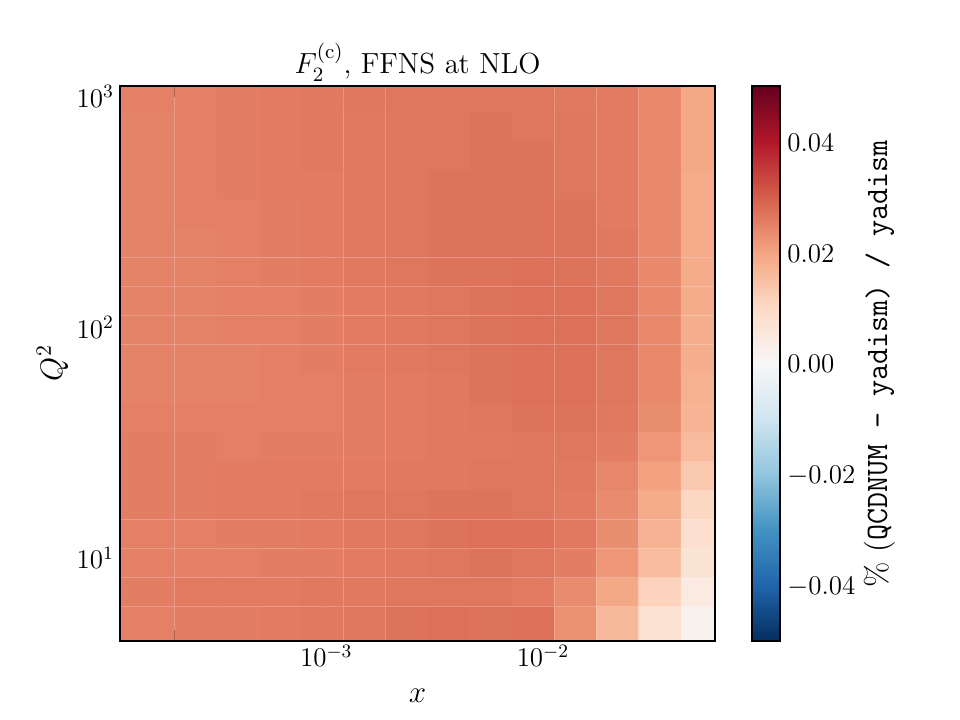}
  \includegraphics[width=.48\textwidth]{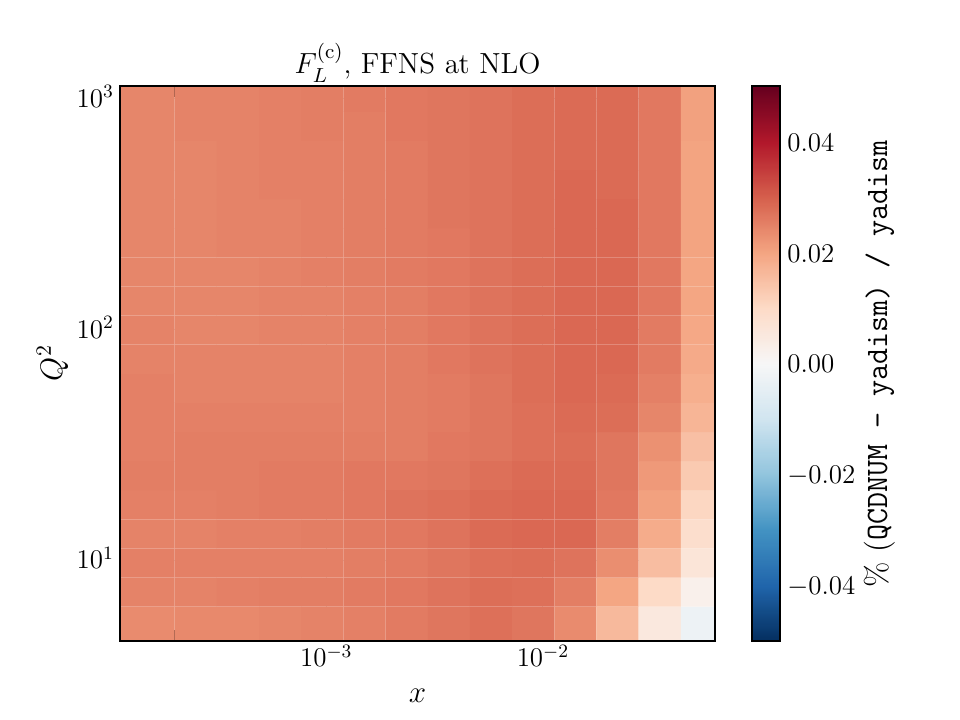}
  \caption{ Similar as \cref{fig:f2_charm_apfelpy} and \cref{fig:fl_charm_apfelpy}, but now comparing
  structure function $F_2^{(c)}$ (left) and $F_L^{(c)}$ (right) computed with \yadism and \qcdnum.
  Note that we can only compare at \nlo due to the
  different conventions adopted by the programs - see text for an explanation.
  }
  \label{fig:f_charm_qcdnum}
\end{figure}

The analogue comparison to \qcdnum is shown in \cref{fig:f_charm_qcdnum} at
\nlo accuracy only demonstrating again a good level of agreement.
Here, we cannot perform the comparison at \nnlo as \qcdnum does not follow
the infrared safe definition of $F^{\text{(c)}}$ (as discussed in \cref{sec:theory}),
but instead includes diagrams with a light quark coupling to the boson into
their respective $F^{(h)}$ result which start contributing at \nnlo.

\subsection{Flavour Number Schemes}
As discussed in \cref{sec:th_options}, \yadism implements various \fns{}s with
the aim of reducing the impact of missing logarithmic or power-like corrections
that become large in certain regions of phase space.
In this section we investigate their relevance.

\begin{figure}
  \includegraphics[width=\textwidth]{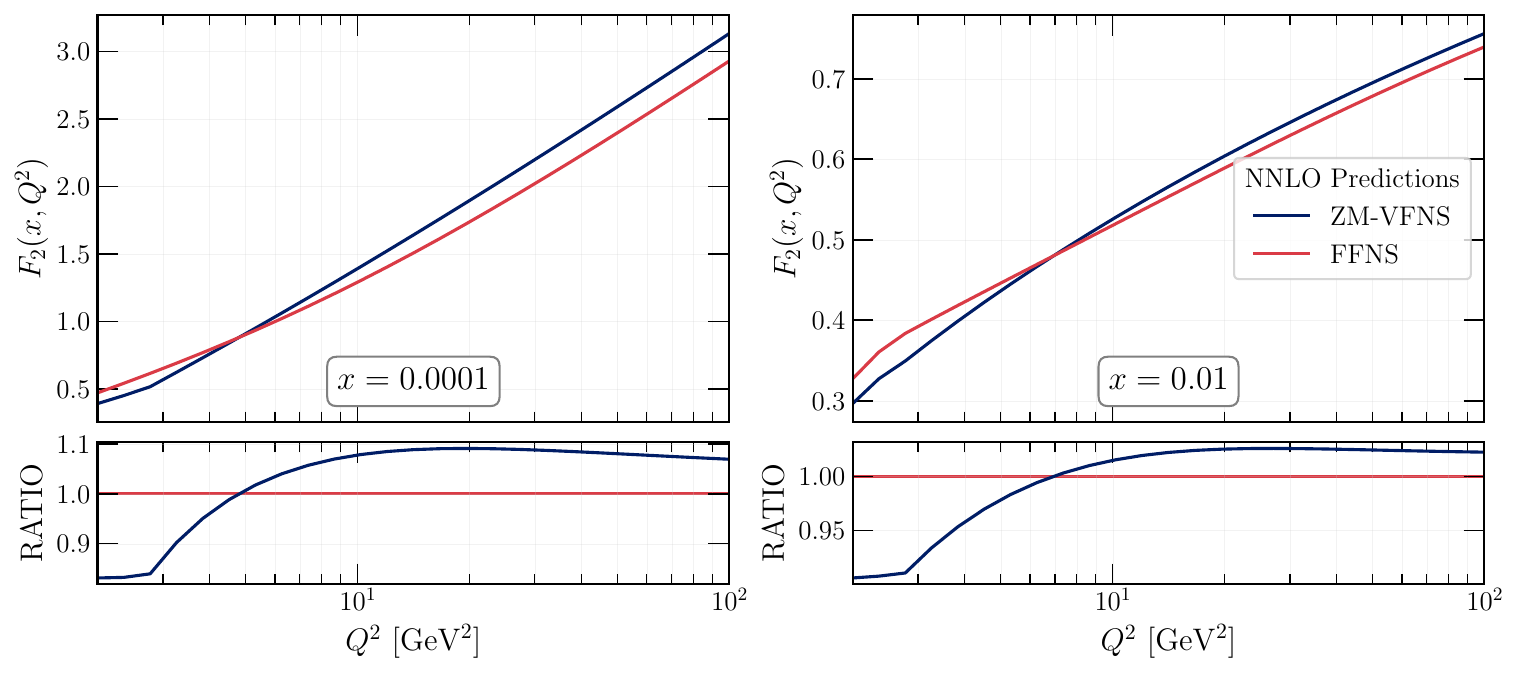}
  \includegraphics[width=\textwidth]{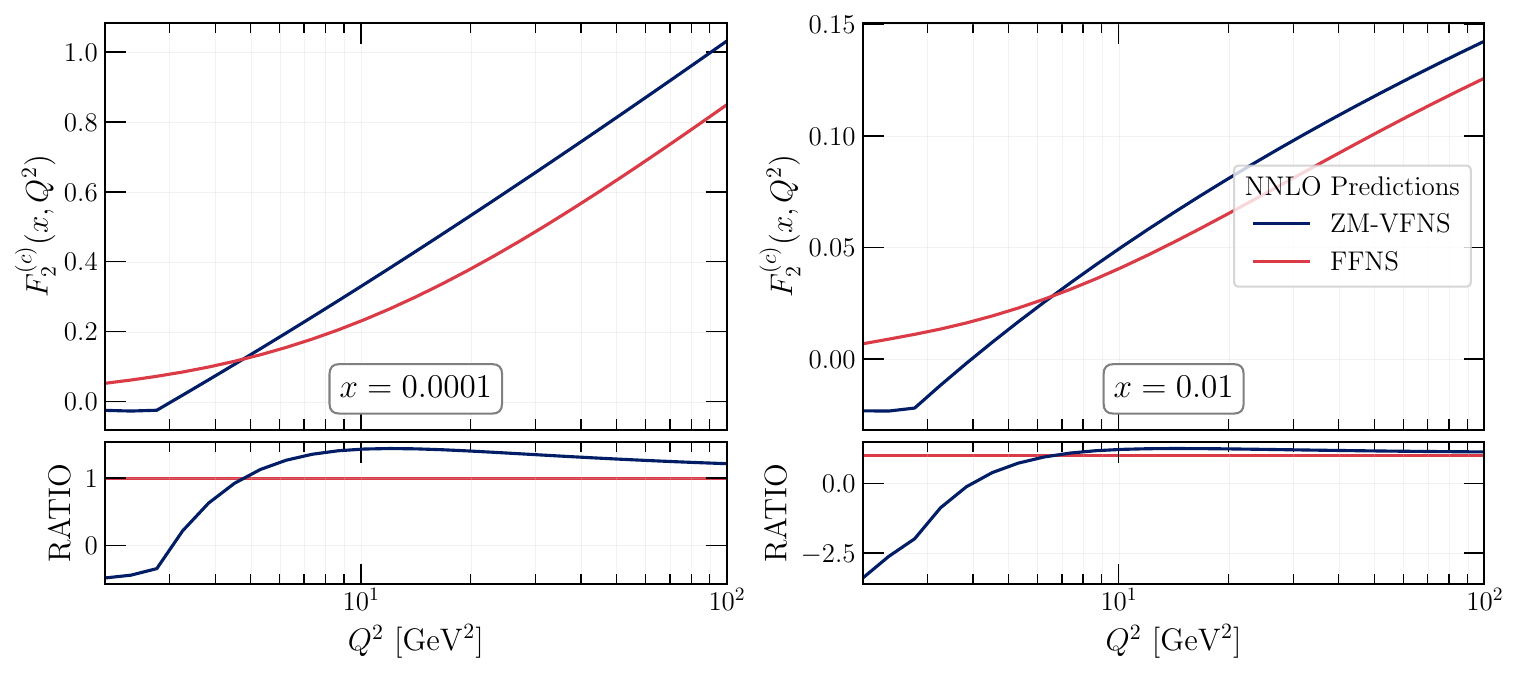}
  \caption{Comparison of the structure functions $F_2$ (top) and $F^{(c)}_2$ (bottom) using \ffns and \zmvfns
  at \nnlo accuracy.
  The top panels show the absolute comparisons while the bottom ones show the ratio w.r.t.\ \zmvfns.
  }
  \label{fig:zm_vs_ffns}
\end{figure}

First, in \cref{fig:zm_vs_ffns}, we compare the \zmvfns and \ffns coefficient functions as a function of $Q^2$.
Recall that the \zmvfns is defined by assuming all (active) quarks to be massless and the \ffns
by considering a single heavy quark with a finite mass and the remaining quarks massless.
We expect both calculations to differ more in the low-$Q^2$ region and
progressively reach better agreement towards the large-$Q^2$ region. However,
while \zmvfns fully resums all (collinear) logarithms $\log(m^2/Q^2)$, \ffns is
a fixed order calculation which only collects a finite number of (collinear)
logarithms and hence a finite difference between the two calculations remains.
We indeed observe for both structure functions $F_2$ and $F_2^{(c)}$ this
expected pattern, thus confirming a consistent implementation.

\begin{figure}
  \includegraphics[width=\textwidth]{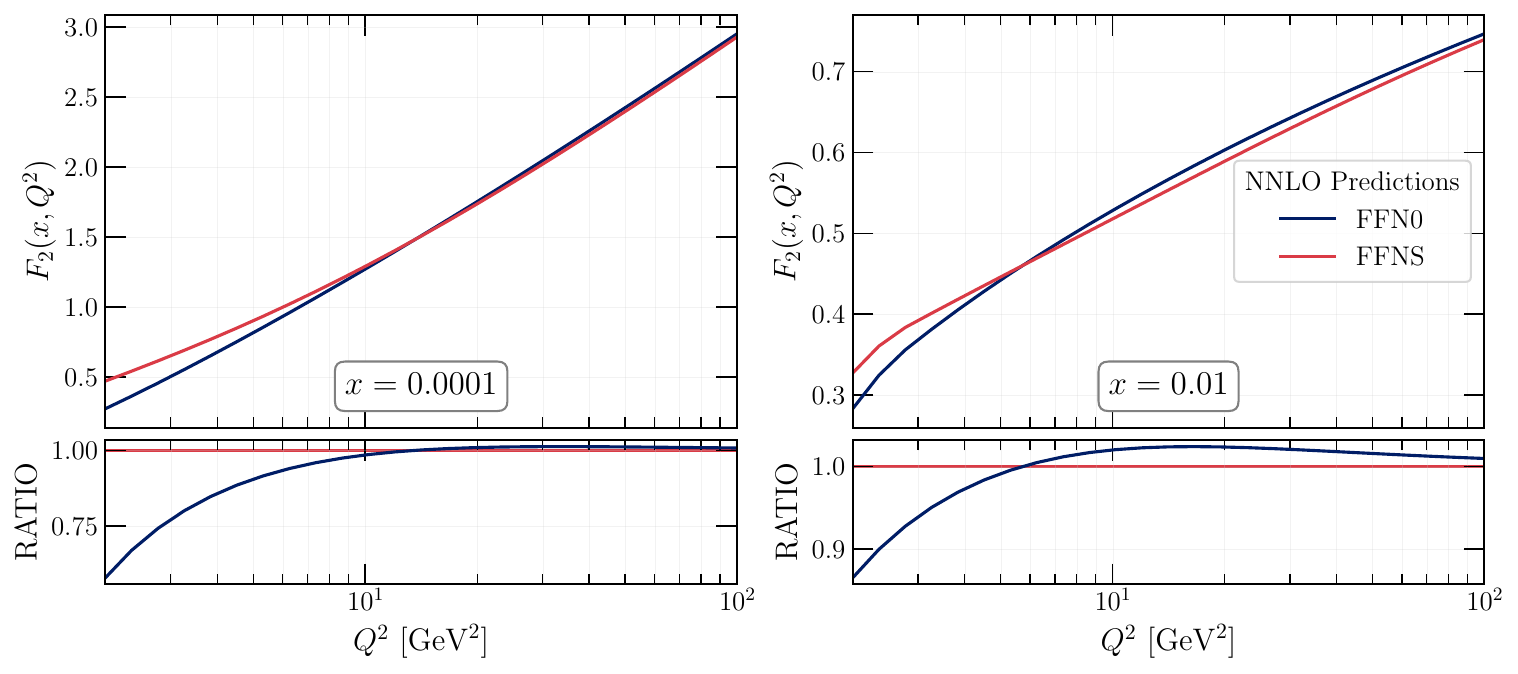}
  \includegraphics[width=\textwidth]{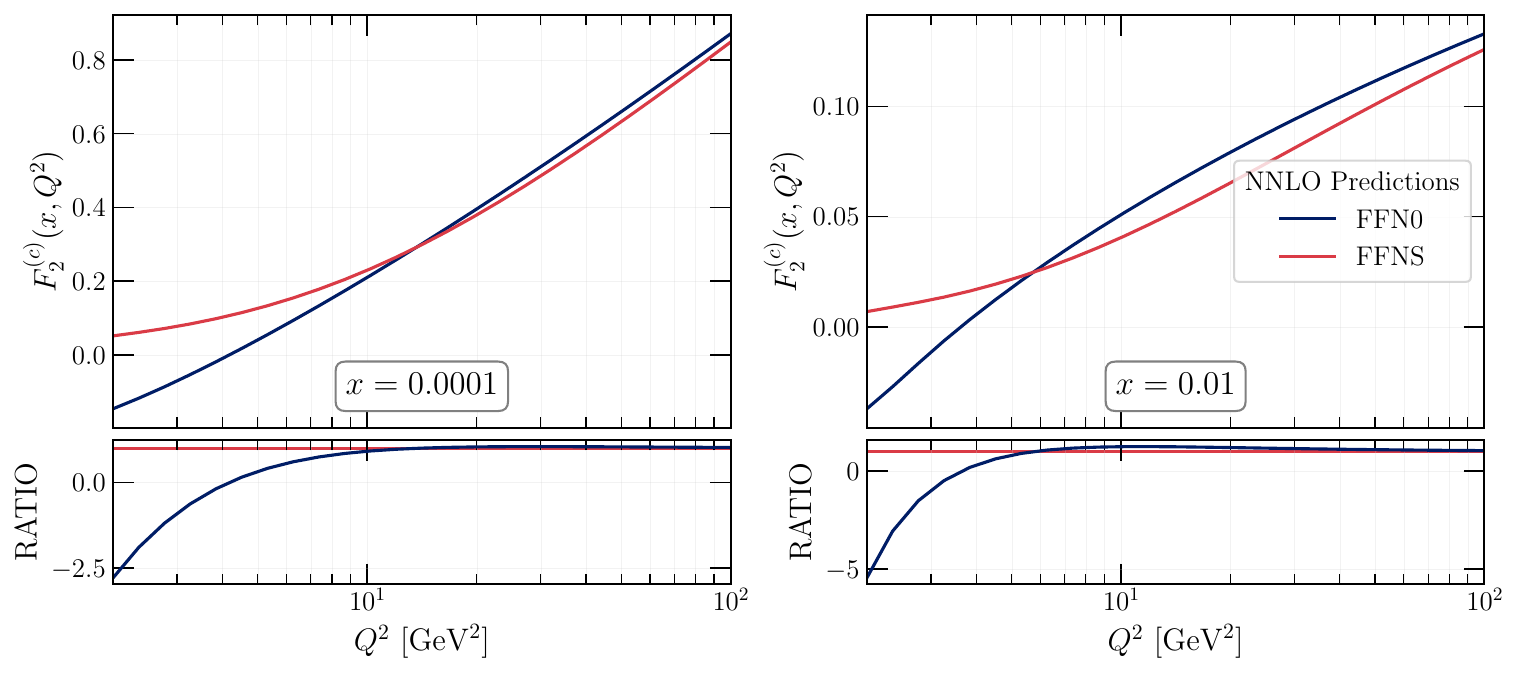}
  \caption{Same as \cref{fig:zm_vs_ffns} but now comparing \ffns and \ffnz.}
  \label{fig:asy_vs_ffns}
\end{figure}

Next, in \cref{fig:asy_vs_ffns}, we compare \ffns and \ffnz coefficient functions as a function of $Q^2$.
Recall that \ffnz is derived from \ffns by only keeping the finite number of collinear logs
and, hence, we expect both calculations to converge in the large-$Q^2$ region where any
power-like mass corrections vanish.
While we can indeed observe this convergence at large-$Q^2$, we also find a
relevant region at mid to low $Q^2$ where mass effects can grow up to $\SI{25}{\percent}$.
This latter region can reach up to $O(100)$ times the heavy quark mass and clearly demonstrates the
need for \gmvfns{}s to improve the accuracy of the prediction.

\section{Conclusions and Outlook}
\label{sec:concl}
In this paper we presented \yadism, a new software package to compute
cross section and structure functions in deep-inelastic scattering. In
\cref{sec:theory} we reviewed some core features that are relevant in specifying
the exact theoretical and experimental setup for which \yadism is able to provides calculations. \Yadism has been developed with much care to ensure the results are in agreement with the widely used packages
\qcdnum and \apfelpp when they should be, and to understand any differences
where they do appear. The success of this effort has been shown in the
benchmarking exercise presented in \cref{sec:bench}.

While \yadism is able to reproduce results also available in \qcdnum and
\apfelpp, it provides value in following a modular design, allowing it to be easily extended with new coefficient functions, new observables or new DIS-like theories
\cite{Bissolotti:2023vdw,McCullough:2022hzr}. At the time of writing, \yadism has
been used for calculations of the photon
PDF~\cite{NNPDF:2024djq}, where the photon PDF is
computed from DIS cross sections in the LuxQED
procedure~\cite{Manohar:2016nzj,Manohar:2017eqh,Bertone:2017bme,Cridge:2021pxm,Xie:2023qbn},
the study of heavy quark mass effects in polarised DIS scatterings~\cite{Hekhorn:2024tqm},
the study of neutrino-ion interactions at the Forward Physics Facility (FPF)~\cite{Cruz-Martinez:2023sdv}
and the determination of low-energy neutrino structure
functions~\cite{Candido:2023utz}. \Yadism is adopted by the NNPDF collaboration to
perform PDF fits where it provides the computations for all fully
inclusive DIS measurements by merit of its interface to \pineappl, and so far this
has resulted in the work presented in Refs.~\cite{NNPDF:2024djq,NNPDF:2024dpb}.

In future it will be possible to adjust the package structure to exploit
synergies with a new software package dedicated to the computation of
cross sections in semi-inclusive annihilation (SIA). Indeed, DIS and SIA are
related by a crossing relation of Feynman diagrams, which makes the mathematical
structure of convoluting a collinear distribution, in this case a fragmentation
function (FF), with a coefficient function very similar.

\acknowledgments
\label{sec:ack}
We thank all the members of the NNPDF collaboration for numerous discussions during the development of this work.
We thank in particular Stefano Forte for guidance during the early stages of the project. We thank Stefano Forte, Juan Rojo, and Christopher Schwan for a careful reading of the manuscript.
We are grateful to Valerio Bertone for discussions about \apfel and \apfelpp, and to Kirill Kudashkin for sharing the results of his NLO CC calculation with heavy initial states.

F.~H.\ is supported by the Academy of Finland
project 358090 and is funded as a part of the Center
of Excellence in Quark Matter of the Academy of Finland, project 346326.
R.~S.\ is supported by the U.K.~Science and Technology Facility Council (STFC) grant ST/T000600/1.
G.~M.\ is supported by NWO, the Dutch Research Council.
T.~R.\ is partially supported by NWO and by the Netherlands eScience Center.

\appendix
\section{NLO CC heavy-to-light coefficient functions}
\label{sec:Kirill}
In this appendix we sketch the setup used in the computation of
\nlo heavy-to-light CC coefficient functions.
For the NC counterpart, the problem has been solved in \cite{kretzer-schienbein}.
There, the authors study the process
\begin{equation}
    Q_1(k_1) + B^*(q) \to Q'(k_2) + X
    \label{eq:nc_ic}
\end{equation}
with heavy quarks $Q$ and $Q'$ and their respective masses $k_1^2=m_1^2$ and $k_2^2=m_2^2$ and the scattered boson $B$ with
virtuality $Q^2 = - q^2$. Eventually they compute the coefficient functions up to NLO,
which agrees with results that were already available (see Ref.~\cite[Sec.~2.2.1]{kretzer-schienbein}).

In our case, for CC DIS, this calculation (beyond the \lo) cannot be used directly, since we are interested in the scattering
\begin{equation}
    Q_1(k_1) + W^*(q) \to q'(k_2) + X
    \label{eq:cc_ic}
\end{equation}
now with $k_2^2 = 0$.
Indeed, in \cref{eq:nc_ic}, the explicit dependency on $m_2^2 > 0$ shields the calculation from additional
infrared singularities, but this does not happen anymore in \cref{eq:cc_ic}.

Thus, the CC coefficient functions, with massive initial states,
require a dedicated calculation and we report the structure of the obtained results.
More details will be discussed in a forthcoming publication~\cite{Kirill}.

In particular, for the intrinsic charm contributions to all three unpolarized structure functions we have:
\begin{align}
    2 F_1(x,Q^2) &\supset \int\limits_x^1\!\frac{d\zeta}{\zeta} C^1_c(\zeta,\alpha_s(Q^2),Q^2/m_c^2) f_c(x/\zeta)\\
    \frac 1 x F_2(x,Q^2) &\supset \int\limits_x^1\!\frac{d\zeta}{\zeta} C^2_c(\zeta,\alpha_s(Q^2),Q^2/m_c^2) f_c(x/\zeta)\\
    F_3(x,Q^2) &\supset \int\limits_x^1\!\frac{d\zeta}{\zeta} C^3_c(\zeta,\alpha_s(Q^2),Q^2/m_c^2) f_c(x/\zeta)
\end{align}
where the intrinsic charm \pdf $f_c(z)$ is now scale independent.
The respective coefficient functions are expanded in powers of $\alpha_s$ as
\begin{align}
    C^1_c(\zeta,\alpha_s(Q^2),Q^2/m_c^2) &= e^2\left(\delta(1-\zeta) + \frac{\alpha_s(Q^2) C_F}{2\pi}C^{1,(1)}_c(\zeta,y) \right)\\
    C^2_c(\zeta,\alpha_s(Q^2),Q^2/m_c^2) &= e^2\frac{(y-1)}{y}\left(\delta(1-\zeta) + \frac{\alpha_s(Q^2) C_F}{2\pi}C^{2,(1)}_c(\zeta,y) \right)\\
    C^3_c(\zeta,\alpha_s(Q^2),Q^2/m_c^2) &= e^2\left(\delta(1-\zeta) + \frac{\alpha_s(Q^2) C_F}{2\pi}C^{3,(1)}_c(\zeta,y) \right)
\end{align}
with $C_F$ the second Casimir constant of the fundamental color representation and $y=-Q^2/m_c^2$.
The lengthy expressions of the actual \nlo coefficient functions $C_c^{k,(1)}$ are provided as ancillary
Mathematica files attached to the arXiv version of this publication.
The reader can note the explicit appearance
of soft-collinear contributions, which manifest themselves as $\left(\frac{\ln(1-\zeta)}{1-\zeta}\right)_+$ and which are not present in
\cite{kretzer-schienbein}.

\section{User Manual}
\label{sec:user-manual}
The purpose of this appendix is to summarize the steps required to generate an interpolation grid using \yadism by providing an explicit example.
For more details, we refer to the online documentation containing, among other things, instructions on how to install and use \yadism:
\begin{center}
  \url{https://yadism.readthedocs.io/en/latest/index.html}
\end{center}

\subsection{Installation}
The requirements to run and install \yadism are a system with \texttt{Python3}\footnote{The range of minor versions of \texttt{Python} supported by a given version of \yadism can be found on the Python Package Index: \url{https://pypi.org/project/yadism}.}, along with the python package installer \texttt{pip} which enables the installation of packages from the Python Package Index.

With these requirements met, \yadism, specifically v0.13.2 discussed in this paper, can be installed by simply running
\begin{lstlisting}[language=Bash]
  python -m pip install yadism==0.13.2
\end{lstlisting}

\subsection{Generating results with \yadism}
Once \yadism is installed, results can be produced by running a single function that takes as input two dictionaries:
one with instructions on the observable to be computed, and one containing the theory parameters of the calculation.
Once the computation has been run, the output can be saved as a \pineappl interpolation grid, or
convoluted directly to any PDF in LHAPDF format and thereby obtain the requested predictions.

At a first glance, the required number of theory parameters might seems
redundant, but this is actually a design choice.
In fact, for most of these parameters, \yadism does not provide a default value,
as we want the user to be fully aware of all the settings entering in the calculation.

The code snippet below provides a simple example of a script that can be used to compute
the reduced charm HERA NC cross section.

\begin{lstlisting}[language=Python]
import lhapdf
import yadism
from yadbox.export import dump_pineappl_to_file

# OBSERVABLE RUNCARD:
observablecard = {
  # Process type: "EM", "NC", "CC"
  "prDIS": "NC",
  # Projectile: "electron", "positron", "neutrino", "antineutrino"
  "ProjectileDIS": "electron",
  # Scattering target: "proton", "neutron", "isoscalar", "lead", "iron", "neon" or "marble"
  "TargetDIS": "proton",
  # Interpolation: if True use log interpolation
  "interpolation_is_log": True,
  # Interpolation: polynomial degree, 1 = linear, ...
  "interpolation_polynomial_degree": 4,
  # Interpolation: xgrid values
  # Note: for illustrative purposes the grid is chosen very small here
  "interpolation_xgrid": [1e-7, 1e-6, 1e-5, 1e-4, 1e-3, 1e-2, 1e-1, 1.0],
  # Observables configurations
  "observables": {
    "XSHERANCAVG_charm": [
      {
          "y": 0.8240707777909629,
          "x": 3e-05,
          "Q2": 2.5,
      },
      {
          "y": 0.3531731904818413,
          "x": 7e-05,
          "Q2": 2.5,
      },
      # Add here the kinematics of other datapoints
    ],
    # Potentially include observables other than XSHERANCAVG_charm,
    # each of them has to be: TYPE_heaviness, where heaviness can take:
    # "charm", "bottom", "top", "total" or "light".
  },
  # Projectile polarization faction, float from 0 to 1.
  "PolarizationDIS": 0.0,
  # Exchanged boson propagator correction
  "PropagatorCorrection": 0.0,
  # Restrict boson coupling to a single parton ? Monte Carlo PID or None for all partons
  "NCPositivityCharge": None,
}

# THEORY RUNCARD:
theorycard = {
  "CKM": "0.97428 0.22530 0.003470 0.22520 0.97345 0.041000 0.00862 0.04030 0.999152", # CKM matrix elements
  "FNS": "FFNS", # Flavour Number Scheme, options: "FFNS", "FFN0", "ZM-VFNS"
  "GF": 1.1663787e-05, # [GeV^-2] Fermi coupling constant
  "IC": 1, # 0 = perturbative charm only, 1 = intrinsic charm allowed
  "MP": 0.938, # [GeV] proton mass
  "MW": 80.398, # [GeV] W boson mass
  "MZ": 91.1876, # [GeV] Z boson mass
  "NfFF": 4, # (fixed) number of running flavors, only for FFNS or FFN0 schemes
  "PTO": 2, # perturbative order in alpha_s: 0 = LO (alpha_s^0), 1 = NLO (alpha_s^1) ...
  "Q0": 1.65, # [GeV] reference scale for the flavor patch determination
  "nf0": 4, # number of active flavors at the Q0 reference scale
  "Qref": 91.2, # [GeV] reference scale for the alphas value
  "nfref": 5, # number of active flavors at the reference scale Qref
  "alphas": 0.118, # alphas value at the reference scale
  "TMC": 1, # include target mass corrections: 0 = disabled, 1 = leading twist, 2 = higher twist approximated, 3 = higher twist exact
  "XIF": 1.0, # ratio of factorization scale over the hard scattering scale
  "XIR": 1.0, # ratio of renormalization scale over the hard scattering scale
  "alphaqed": 0.007496252, # alpha_em value
  "kcThr": 1.0, # ratio of the charm matching scale over the charm mass
  "kbThr": 1.0, # ratio of the bottom matching scale over the bottom mass
  "ktThr": 1.0, # ratio of the top matching scale over the top mass
  "mc": 1.51, # [GeV] charm mass
  "mb": 4.92, # [GeV] bottom mass
  "mt": 172.5, # [GeV] top mass
  "n3lo_cf_variation": 0, # N3LO coefficient functions variation: -1 = lower bound, 0 = central , 1 = upper bound
  "QED": 0, # QED correction to running of strong coupling: 0 = disabled, 1 = allowed
  "MaxNfAs": 5, # maximum number of flavors in running of strong coupling
  "HQ": "POLE", # heavy quark mass scheme (not yet implemented in yadism)
  "MaxNfPdf": 5, # maximum number of flavors in running of PDFs (ignored by yadism)
  "ModEv": "EXA", # evolution solver for PDFs (ignored by yadism)
}

out = yadism.run_yadism(theorycard, observablecard)

# After running yadism one can either store the result in a pineappl grid
dump_pineappl_to_file(out,"outputgrid.pineappl.lz4","XSHERANCAVG_charm")

# ... or apply a chosen PDF to compute observable predictions direcly
pdf = lhapdf.mkPDF("NNPDF40_nnlo_as_01180")
values = out.apply_pdf(pdf)
\end{lstlisting}

\bibliographystyle{styles/JHEP}
\bibliography{
	bibliography/refs.bib,
	bibliography/exp.bib
}

\listoffixmes

\end{document}